\documentclass[longauth]{aa} % for the long lists of affiliations 
\usepackage{graphicx}	% Including Fig.~files
\usepackage{amsmath}	% Advanced maths commands
\usepackage{multirow}
\usepackage{subfigure}
\usepackage{adjustbox}
\usepackage{xcolor}
\usepackage{float}
\usepackage{amsmath}	% Advanced maths commands
\usepackage{lscape} 
\usepackage{caption}
\usepackage[switch]{lineno}
\usepackage{txfonts}
\usepackage{natbib}
\usepackage[colorlinks=true,linkcolor=blue,allcolors=blue]{hyperref}
\bibpunct{(}{)}{;}{a}{}{,}             %% natbib format for A&A and ApJ

\newcommand{\orcit}[1]{\protect\href{https://orcid.org/#1}{\protect\includegraphics[width=8pt]{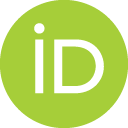}}}

\begin{document}

   \title{Optical variability of the blazar 3C 371: From minute to year timescales}

   \author{J. Otero-Santos
          \orcit{0000-0002-4241-5875}\inst{\ref{csic} \and \ref{iac}}\fnmsep\thanks{Contact: \email{joteros@iaa.es}}
          \and
          C. M. Raiteri\orcit{0000-0003-1784-2784}\inst{\ref{oato}}
          \and
          J. A. Acosta-Pulido\orcit{0000-0002-0433-9656}\inst{\ref{iac} \and \ref{ull}}
          \and
          M. I. Carnerero\orcit{0000-0001-5843-5515}\inst{\ref{oato}}
          \and
          M. Villata\orcit{0000-0003-1743-6946}\inst{\ref{oato}}
          \and
          S. S. Savchenko\orcit{0000-0003-4147-3851}\inst{\ref{stp} \and \ref{sao} \and \ref{pulkovo}}
          \and
          D. Carosati\inst{\ref{ept} \and \ref{inaf-tng}}
          \and
          W. P. Chen\orcit{0000-0003-0262-272X}\inst{\ref{taiwan}}
          \and
          S. O. Kurtanidze\orcit{0000-0002-0319-5873}\inst{\ref{abastumani}}
          \and
          M. D. Joner\orcit{0000-0003-0634-8449}\inst{\ref{wmo}}
          \and
          E. Semkov\orcit{0000-0002-1839-3936}\inst{\ref{sofia_obs}}
          \and
          T. Pursimo\inst{\ref{not} \and \ref{aarhus}}
          \and
          E. Benítez\orcit{0000-0003-1018-2613}\inst{\ref{unam}}
          \and
          G. Damljanovic\orcit{0000-0002-6710-6868}\inst{\ref{vidojevica}}
          \and 
          G. Apolonio\inst{\ref{wmo}}
          \and
          G. A. Borman\orcit{0000-0002-7262-6710}\inst{\ref{CrAO}}
          \and
          V. Bozhilov\inst{\ref{sofia_univ}}
          \and
          F. J. Galindo-Guil\orcit{0000-0003-4776-9098}\inst{\ref{cefca}}
          \and
          T. S. Grishina\orcit{0000-0002-3953-6676}\inst{\ref{stp}}
          \and
          V. A. Hagen-Thorn\orcit{0000-0002-6431-8590}\inst{\ref{stp}}
          \and
          D. Hiriart\orcit{0000-0002-4711-7658}\inst{\ref{unam2}}
          \and
          H. Y. Hsiao\inst{\ref{taiwan}}
          \and
          S. Ibryamov\orcit{0000-0002-4618-1201}\inst{\ref{shumen}}
          \and
          R. Z. Ivanidze\inst{\ref{abastumani}}
          \and
          G. N. Kimeridze\inst{\ref{abastumani}}
          \and
          E. N. Kopatskaya\orcit{0000-0001-9518-337X}\inst{\ref{stp}}
          \and
          O. M. Kurtanidze\orcit{0000-0001-5385-0576}\inst{\ref{abastumani} \and \ref{engelhardt}}
          \and
          V. M. Larionov\inst{\ref{stp}}
          \and
          E. G. Larionova\orcit{0000-0002-2471-6500}\inst{\ref{stp}}
          \and
          L. V. Larionova\orcit{0000-0002-0274-1481}\inst{\ref{stp}}
          \and
          M. Minev\inst{\ref{sofia_univ} \and \ref{sofia_obs}}
          \and
          D. A. Morozova\orcit{0000-0002-9407-7804}\inst{\ref{stp}}
          \and
          M. G. Nikolashvili\orcit{0000-0003-0408-7177}\inst{\ref{abastumani}}
          \and
          E. Ovcharov\inst{\ref{sofia_univ}}
          \and
          L. A. Sigua\orcit{0000-0002-6410-1084}\inst{\ref{abastumani}}
          \and
          M. Stojanovic\orcit{0000-0002-4105-7113}\inst{\ref{vidojevica}}
          \and
          I. S. Troitskiy\orcit{0000-0002-4218-0148}\inst{\ref{stp}}
          \and
          Yu. V. Troitskaya\orcit{0000-0002-9907-9876}\inst{\ref{stp}}
          \and
          A. Tsai\inst{\ref{taiwan} \and \ref{sunyat-se}}
          \and
          A. Valcheva\inst{\ref{sofia_univ}}
          \and
          A. A. Vasilyev\orcit{0000-0002-8293-0214}\inst{\ref{stp}}
          \and
          O. Vince\orcit{0009-0008-5761-3701}\inst{\ref{vidojevica}}
          \and
          E. Zaharieva\inst{\ref{sofia_univ}}
          \and
          A. V. Zhovtan\inst{\ref{CrAO}}
          }

   \institute{Instituto de Astrof\'isica de Andalucía (CSIC), Glorieta de la Astronomía s/n, 18008 Granada, Spain \label{csic}
         \and
             Instituto de Astrof\'isica de Canarias (IAC), E-38200 La Laguna, Tenerife, Spain \label{iac}
         \and
             INAF-Osservatorio Astrofisico di Torino, via Osservatorio 20, 10025 Pino Torinese, Italy \label{oato}
         \and  
             Universidad de La Laguna (ULL), Departamento de Astrof\'isica, E-38206 La Laguna, Tenerife, Spain \label{ull}
         \and
             Saint Petersburg State University, 7/9 Universitetskaya nab., St. Petersburg, 199034 Russia \label{stp}
         \and 
             Special Astrophysical Observatory, Russian Academy of Sciences, 369167, Nizhnii Arkhyz, Russia	\label{sao}
         \and
             Pulkovo Observatory, St. Petersburg, 196140, Russia \label{pulkovo}
         \and
             EPT Observatories, Tijarafe, La Palma, Spain \label{ept}
         \and
             INAF, TNG Fundaci\'on Galileo Galilei, La Palma, Spain \label{inaf-tng}
         \and
             Institute of Astronomy, National Central University, Taoyuan 32001, Taiwan \label{taiwan}
         \and
             Abastumani Observatory, Mt. Kanobili, 0301 Abastumani, Georgia \label{abastumani}
         \and
             Department of Physics and Astronomy, N283 ESC, Brigham Young University, Provo, UT 84602, USA \label{wmo}
         \and
             Institute of Astronomy and National Astronomical Observatory, Bulgarian Academy of Sciences, 72 Tsarigradsko shosse Blvd., 1784 Sofia, Bulgaria \label{sofia_obs}
         \and
             Nordic Optical Telescope, Apartado 474, E-38700 Santa Cruz de La Palma, Santa Cruz de Tenerife, Spain \label{not}
         \and 
             Department of Physics and Astronomy, Aarhus University, Munkegade 120, DK-8000 Aarhus C, Denmark \label{aarhus}
         \and
             Universidad Nacional Aut\'onoma de M\'exico, Instituto de Astronom\'ia, AP 70-264, CDMX 04510, Mexico \label{unam}
         \and
             Astronomical Observatory, Volgina 7, 11060 Belgrade, Serbia \label{vidojevica} 
         \and
             Crimean Astrophysical Observatory RAS, P/O Nauchny, 298409, Russia \label{CrAO}
         \and 
             Department of Astronomy, Faculty of Physics, Sofia University ``St. Kliment Ohridski'', 5 James Bourchier Blvd., BG-1164 Sofia, Bulgaria \label{sofia_univ}
         \and
             Centro de Estudios de Física del Cosmos de Aragón (CEFCA), Plaza San Juan 1, 44001 Teruel, Spain \label{cefca}
         \and
             Universidad Nacional Aut\'onoma de M\'exico, Instituto de Astronom\'ia, AP 106, Ensenada 22800, Baja California, Mexico \label{unam2}
         \and
             Department of Physics and Astronomy, Faculty of Natural Sciences, University of Shumen, 115, Universitetska Str., 9712 Shumen, Bulgaria \label{shumen}
         \and 
             Engelhardt Astronomical Observatory, Kazan Federal University, Tatarstan, Russia \label{engelhardt}
         \and 
             National Sun Yat-sen University, No. 70 Lien-hai Road, Kaohsiung, Taiwan 804201 \label{sunyat-se}
             } 

   \date{Received September 15, 1996; accepted March 16, 1997}

% \abstract{}{}{}{}{} 
% 5 {} token are mandatory
 
  \abstract
   {The BL Lac object 3C~371 was observed by the Transiting Exoplanet Survey Satellite (TESS) for approximately a year, between July 2019 and July 2020, with an unmatched two-minute imaging cadence. In parallel, the Whole Earth Blazar Telescope (WEBT) Collaboration organized an extensive observing campaign, providing three years of continuous optical monitoring between 2018 and 2020. These datasets allow for a thorough investigation of the variability of the source.}
   {The goal of this study is to evaluate the optical variability of 3C~371. Taking advantage of the remarkable cadence of TESS data, we aim to characterize the intra-day variability (IDV) displayed by the source and identify its shortest variability timescale. With this estimate, constraints on the size of the emitting region and black hole mass can be calculated. Moreover, WEBT data are used to investigate long-term variability (LTV), including in terms of the spectral behavior of the source and the polarization variability. Based on the derived characteristics, we aim to extract information on the origin of the variability on different timescales.}
   {We evaluated the variability of 3C 371 by applying the variability amplitude tool, which quantifies variability of the emission. Moreover, we employed common tools, such as ANOVA (ANalysis Of VAariance) tests, wavelet and power spectral density (PSD) analyses to characterize the shortest variability timescales present in the emission and the underlying noise affecting the data. We evaluated the short- and long-term color behavior to understand to understand its spectral behavior. The polarized emission was analyzed, studying its variability and possible rotation patterns of the electric vector position angle (EVPA). Flux distributions of the IDV and LTV were also studied with the aim being to link the flux variations to turbulent and/or accretion-disk-related processes.}
   {Our ANOVA and wavelet analyses reveal several entangled variability timescales. We observe a clear increase in the variability amplitude with increasing width of the time intervals evaluated. We are also able to resolve significant variations on timescales of as little as $\sim$0.5~hours. The PSD analysis reveals a red-noise spectrum with a break at IDV timescales. The spectral analysis shows a mild bluer-when-brighter (BWB) trend on long timescales. On short timescales, mixed BWB, achromatic and redder-when-brighter (RWB) signatures can be observed. The polarized emission shows an interesting slow EVPA rotation during the flaring period, {where a simple stochastic model can be excluded as the origin with a 3$\sigma$ significance}. The flux distributions show a preference for a Gaussian model for the IDV, and suggest it may be linked to turbulent processes, while the LTV is better represented by a log-normal distribution and may have a disk-related.}
   {}


   \maketitle
%
%-------------------------------------------------------------------

\section{Introduction}

Blazars, a subtype of radio-loud active galactic nuclei (AGN), develop relativistic jets that are closely aligned with the line of sight, which leads to a relativistic boosting of their emission. They can be divided {into} BL Lacertae (BL Lac) objects and flat-spectrum radio quasars (FSRQs) depending on the properties of their optical spectrum \citep{urry1995}. BL Lacs typically present an (almost) featureless optical spectrum, while FSRQs show broad emission lines with an equivalent width $|EW|>5$~\AA~in their rest frame \citep{stickel1991}. Their broadband emission is mainly non thermal, is associated to the relativistic jet, and can extend from radio frequencies up to $\gamma$-ray energies. Their spectral energy distribution (SED) shows a typical double-bump structure \citep[see e.g.,][]{abdo2010}. The low-energy bump is produced by synchrotron radiation of the electrons moving under the influence of the magnetic field of the jet \citep[see e.g.,][]{koenigl1981}. The high-energy bump is commonly modeled through inverse Compton (IC) scattering of low-energy photons with the same population of relativistic electrons, that is, within the leptonic interpretation. This can happen with the same low-energy photons of the synchrotron radiation through synchrotron self-Compton scattering \citep[SSC; see][]{maraschi1992}, or with low-energy photons from outside the jet --- if there is an external injection of photons --- through external Compton scattering \citep[EC; see][]{dermer1993}. While this is the commonly adopted explanation, hadronic models have also been used in recent years \citep[see e.g.,][]{cerruti2015}.

Variability is a key signature of the blazar emission. This variability is present in their broadband emission, manifesting on all timescales \citep{fan2005,singh2020}. Variations in their emission on timescales of several months or years are a typical feature of the emission of blazars, and this is known as long-term variability \citep[LTV; see e.g.,][]{rajput2020,bhatta2021}. Faster variations are also often observed, that is, on timescales of several days or weeks, as short-term variability \citep[STV; see][]{gupta2008,rani2010}. In addition, in some cases blazars have shown remarkably fast variability, on timescales of shorter than one day, down to even a few minutes, this is intraday variability \citep[IDV; e.g.,][]{wagner1995,villata2002,raiteri2021,raiteri2021b}. Different timescales have been associated with different processes and emission mechanisms. However, variability patterns on different timescales are typically entangled and very complex, resulting in a challenging physical interpretation. Nevertheless, variability has become a major tool for understanding the physics taking place in the relativistic jets of blazars. 

3C~371 is a BL Lac object located at a distance of $z=0.0510 \pm 0.0003$, as estimated from optical spectroscopic observations by \cite{degrijp1992}. \cite{miller1975} also classified this source as an intermediate case between the BL Lac type and a radio galaxy. 3C~371 is hosted by a bright galaxy with $R=14.22\pm 0.03$ \citep{purismo2002,nilsson2003}, which makes an important emission contribution in the optical and near-infrared bands \citep{raiteri2014}. It has been reported to show intense variability in its broadband emission on a wide range of timescales: several years in radio \citep{nieppola2009} and ultraviolet (UV) wavelengths \citep{paltani1994}; {changes of more than 1 magnitude on timescales of years and smaller variations in scales of several days in the optical} \citep[see][]{oke1967}; or long-term variations in X-ray observations performed by EXOSAT \citep{giommi1990}. It has also been cataloged as variable in the $\gamma$-ray regime, as shown by the results presented in the \textit{Fermi}-LAT Fourth Source Catalog Data Release 3 (4FGL-DR3) catalog, where a variability index of $\sim$133 is reported \citep{abdollahi2020,abdollahi2022}. Moreover, faster variability has also been observed for 3C~371. For instance, \cite{kraus2003} report this source as a fast variable radio source, with variability on timescales of $\leq$2days which is, similar to the 2.4 day timescales observed by \cite{heidt1996} in the optical $R$-band. However, interstellar scintillation cannot be ruled out as the origin of such fast variability in the radio band. It has also been reported to show IDV in the optical band by \cite{xilouris2006}, making this object an extremely interesting target for performing variability studies down to the shortest variability timescales.

These interesting variability patterns led to an extensive monitoring of 3C~371 by the Whole Earth Blazar Telescope\footnote{\url{https://www.oato.inaf.it/blazars/webt}} (WEBT) Collaboration in optical and radio frequencies. A proposal to the Transiting Exoplanet Survey Satellite (TESS) led to observations with an exceptional cadence of 2 minutes between 2019 and 2020. Taking advantage of these extensive datasets, we have conducted a variability study of 3C~371, evaluating the different characteristic variability timescales in its optical emission between 2018 and 2020, when the source was extensively monitored by the WEBT Collaboration.

The present paper is structured as follows: All the optical datasets are detailed in Sect.~\ref{sec2}. We present an evalutaion of the IDV in Sect.~\ref{sec3} that makes use of the TESS data. A variability analysis taking advantage of all the WEBT data collected is presented in Sect.~\ref{sec4}. In Sect.~\ref{sec5} we show how we evaluated the optical spectral behavior during the observed period. A study of the variability of the optical polarized emission is presented in Sect.~\ref{sec_polarization} and a discussion of the possible origin of the observed variability in relation to the measured flux distributions on different timescales is presented in Sect.~\ref{sec7}. Finally, the main conclusions of this work are summarized in Sect.~\ref{sec8}. 

%--------------------------------------------------------------------
\section{Observations and data reduction}\label{sec2}

%-------------------------------------- Two column figure (place early!)
\subsection{TESS photometric data}
\begin{figure}
        \includegraphics[width=\columnwidth]{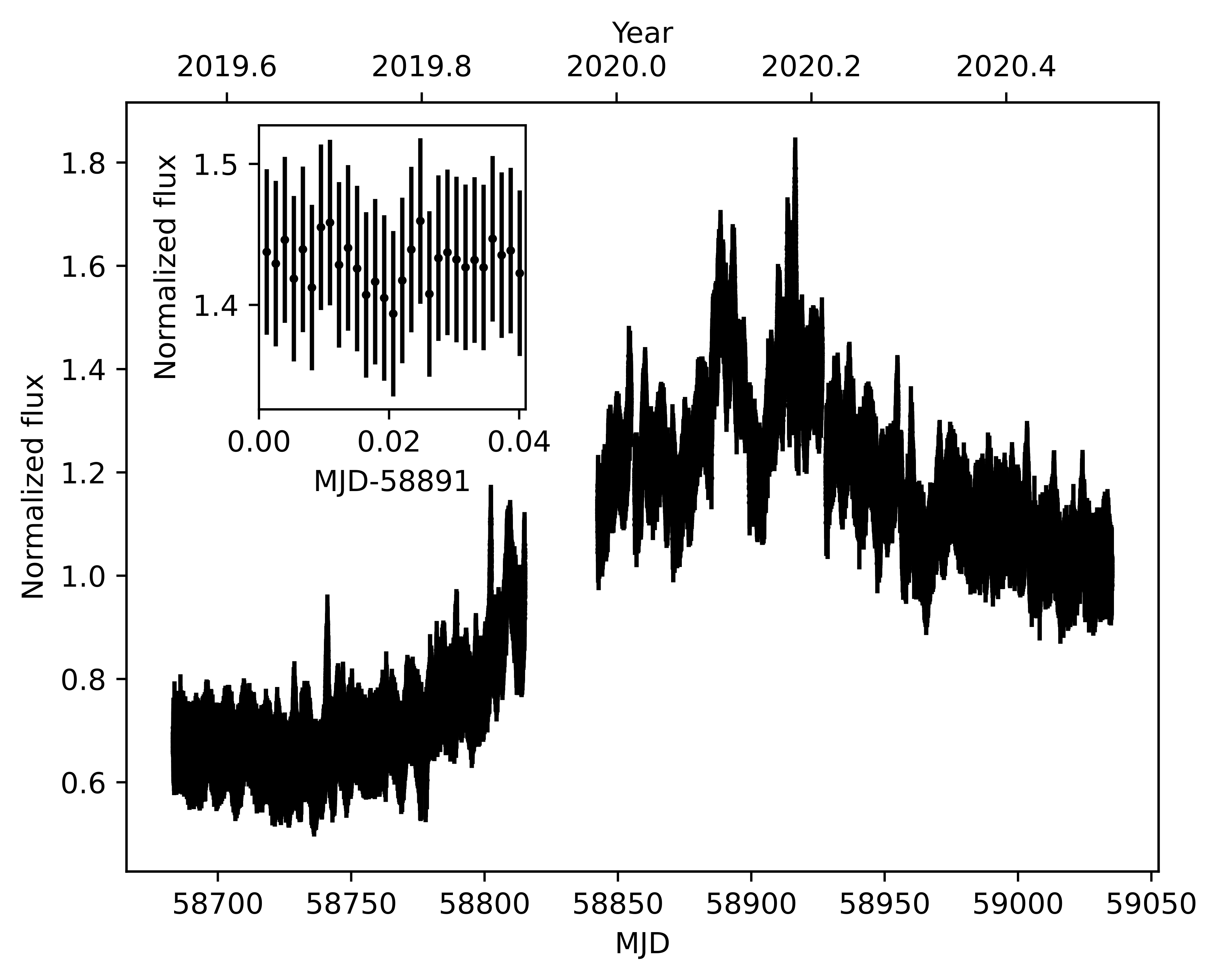}
    \caption{TESS SAP\_FLUX light curve of 3C~371. The offsets between the different sectors were estimated using the observed $R$-band WEBT light curve, and were then converted into arbitrary flux units and normalized by the mean value. The inset panel corresponds to a zoom onto a 1 hour of the light curve to illustrate the cadence and uncertainties of the observations.}
    \label{fig:tess_flux}
\end{figure}

3C 371 has been a target of the TESS mission with an observing cadence of 2 minutes for almost 1 year, with data available during 12 TESS observing sectors during cycle 2, detailed in Table \ref{tab:TESS_sectors}, leading to a total of 214236 data points. These data, downloaded from the Mikulski Archive for Space Telescopes\footnote{\url{https://mast.stsci.edu/portal/Mashup/Clients/Mast/Portal.html}} (MAST), contain information of the simple aperture photometry electron flux (SAP\_FLUX, in electrons per second), that will be used hereafter for the data reduction and analysis presented in this study {\citep{feinstein2019}}. They also contain presearch conditioned simple aperture photometry fluxes (PDCSAP\_FLUX). However, as discussed by \cite{raiteri2021,raiteri2021b}, they are not suitable for blazar variability studies as they remove long-term trends in the data that may be genuine variability signatures of the source. {SAP\_FLUX have a poorer background subtraction. However, owing to the brightness of the source and the good agreement with the WEBT data, we found SAP\_FLUX values to be suitable for this study.} In order to correct for systematic scaling offsets between different sectors, we scaled each sector by a certain offset, calculated from a direct comparison with the WEBT $R$-band light curve (see Sect.~\ref{sec2.2}). Moreover, we removed outliers and points showing an extremely high scattering ($\sim$0.01\% of the points), resulting in a total of 214216 observations included in the light curve. Finally, we also normalized the scaled light curve by the mean SAP\_FLUX value. The light curve of the scaled, normalized aperture photometry flux is shown in Figure \ref{fig:tess_flux}.

\begin{table}
\centering
\caption{Starting and ending dates of each TESS sector during which 3C~371 was observed.}
\label{tab:TESS_sectors}
\resizebox{\columnwidth}{!}{%
\begin{tabular}{ccc}
\hline
\multirow{2}{*}{TESS sector} & Time interval & Time interval \\
                  & [MJD] & [Date] \\ \hline
Sector 14 & 58682 -- 58710 & 18 Jul. 2019 -- 15 Aug. 2019 \\ \hline
Sector 15 & 58711 -- 58737 & 16 Aug. 2019 -- 11 Sep. 2019 \\ \hline
Sector 16 & 58738 -- 58763 & 12 Sep. 2019 -- 07 Oct. 2019 \\ \hline
Sector 17 & 58764 -- 58789 & 08 Oct. 2019 -- 02 Nov. 2019 \\ \hline
Sector 18 & 58790 -- 58814 & 03 Nov. 2019 -- 27 Nov. 2019 \\ \hline
Sector 20 & 58842 -- 58869 & 25 Dec. 2019 -- 21 Jan. 2020 \\ \hline
Sector 21 & 58870 -- 58897 & 22 Jan. 2020 -- 18 Feb. 2020 \\ \hline
Sector 22 & 58899 -- 58926 & 20 Feb. 2020 -- 18 Mar. 2020 \\ \hline
Sector 23 & 58928 -- 58954 & 20 Mar. 2020 -- 15 Apr. 2020 \\ \hline
Sector 24 & 58955 -- 58982 & 16 Apr. 2020 -- 13 May 2020 \\ \hline
Sector 25 & 58983 -- 59009 & 14 May 2020 -- 09 Jun. 2020 \\ \hline
Sector 26 & 59010 -- 59035 & 10 Jun. 2020 -- 05 Jul. 2020 \\ \hline
\end{tabular}
}
\end{table}

\subsection{WEBT optical observations}\label{sec2.2}
\begin{figure}
        \includegraphics[width=\columnwidth]{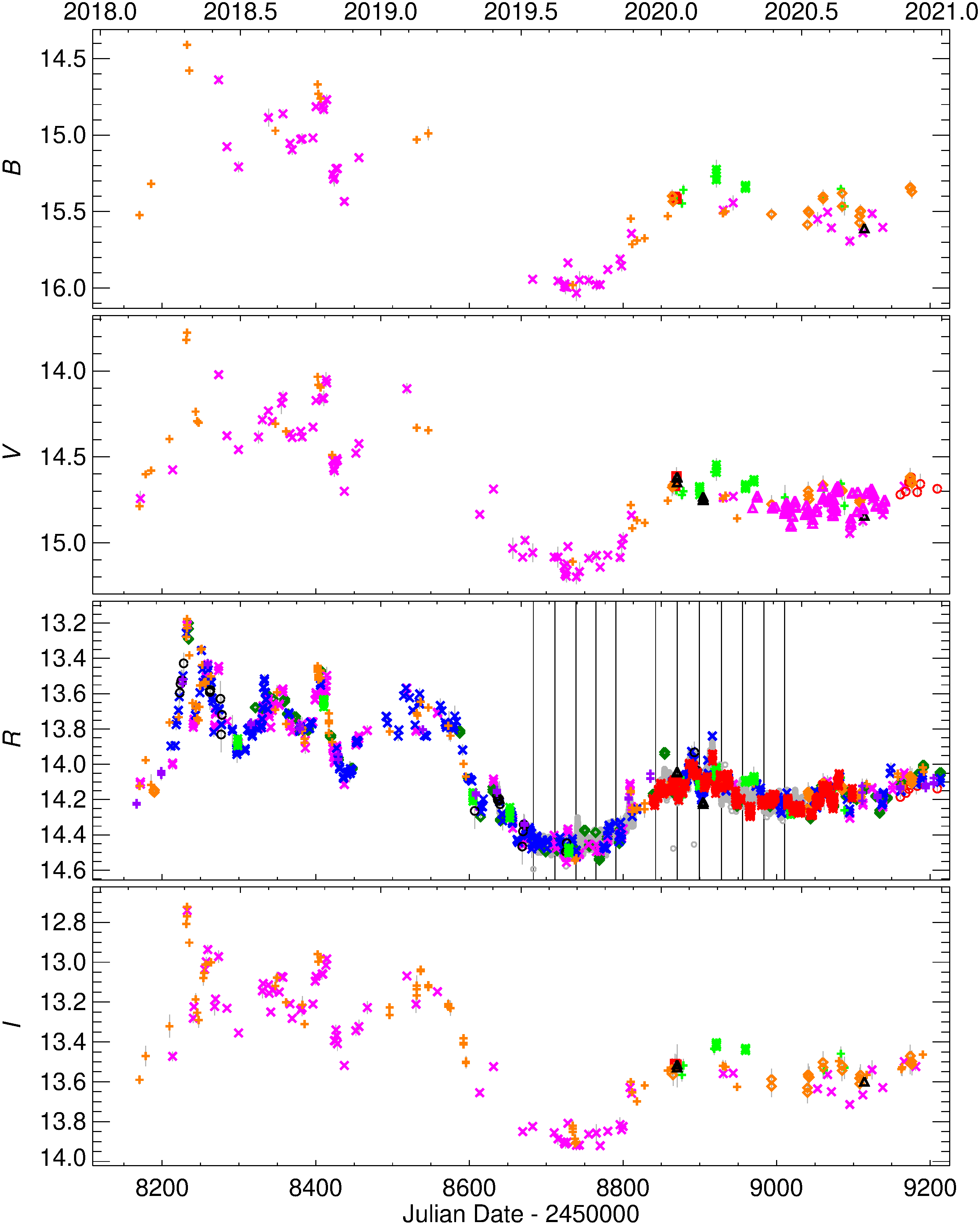}
    \caption{WEBT multiband optical light curves (observed magnitudes) of 3C~371 during the $2018–2020$ observing season. Different colors and symbols are used to distinguish the contributing datasets, as specified in Table~\ref{tab:WEBT_optical_radio_obs}. Vertical lines represent the start of the different TESS sectors during which the source was observed. The $R$-band calibrated TESS data are represented with gray markers in the third panel.}
    \label{fig:webt_LCs}
\end{figure}

This blazar is also a regularly monitored target by the WEBT Collaboration
%\footnote{\url{https://www.oato.inaf.it/blazars/webt/}} 
with several telescopes distributed around the globe. It has been observed in the Johnson--Cousins $BVRI$ bands, with a specially remarkable coverage in the $R$ band. In particular, we focus here in the period between January 2018 and December 2020. The observed optical light curves constructed with the data taken by the WEBT Collaboration are displayed in Fig.~\ref{fig:webt_LCs}, and all the telescopes that provided data are listed in Table \ref{tab:WEBT_optical_radio_obs}. 

\begin{table*}
\centering
\caption{WEBT observatories supporting the observing campaign of 3C~371 in the optical $BVRI$ bands. The telescope size is reported (cm), together with the total number of observations provided, symbol and color used in the light-curve plots.}
\label{tab:WEBT_optical_radio_obs}
\begin{tabular}{llccccc}
\hline
Observatory & Country & Telescope size (cm) & Band & N & Symbol & Color \\ 
\hline
Abastumani & Georgia & 70 & $R$ & 253 & {\LARGE $\diamond$} &  dark green\\ 
Crimean$^{a}$   & Crimea  & 70 & $BVRI$ & 377 & $\times$ &  magenta\\ 
Hans Haffner & Germany & 50 & $VR$ & 16 & {\LARGE $\circ$} & red \\
Lulin & Taiwan & 40 & $R$ & 371 & $\times$ & blue\\
Roque (NOT) & Spain & 260 & $BVRI$ & 29 & {\large $+$} & green\\
Rozhen & Bulgaria & 200 & $BVRI$ & 12 & {$\square$} & red \\
Rozhen & Bulgaria & 50/70 & $BVRI$ & 91 & {\LARGE $\diamond$} & orange\\
San Pedro Martir$^{a}$ & Mexico & 84 & $R$ & 28 & {\LARGE $\circ$} & black\\
St.~Petersburg$^{a}$ & Russia & 40 & $BVRI$ & 198 & {\large $+$} & orange\\
Teide (IAC80) & Spain & 80 & $BVRI$& 119 & {\LARGE $\ast$} & green\\
Teide (STELLA-I) & Spain & 120 & $R$ & 84 & {\large $+$}& violet\\
Tijarafe & Spain & 40 & $R$ & 466 & {\LARGE $\ast$} & red\\
Vidojevica$^b$ & Serbia & 60 & $BVRI$ & 18 & {$\triangle$} & black\\
West Mountain & US & 91 & $V$ & 112 & {$\triangle$} & magenta\\
\hline
\end{tabular}\\
{\flushleft
\vspace{-0.2cm}
\hspace{2.6cm}$^a$Telescopes with polarimetric instrumentation available.\\
\hspace{2.6cm}$^b$Astronomical Station Vidojevica.\\
}
\end{table*}

\begin{figure}
    \includegraphics[width=\columnwidth]{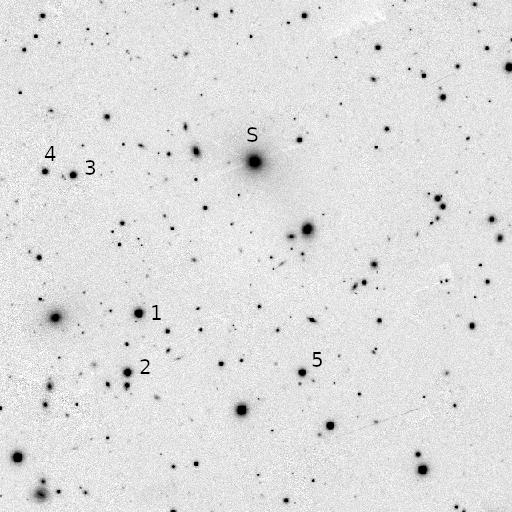}
    \caption{Finding chart obtained from the PanSTARRS-1 (PS1) Image Cutout Service\protect\footnotemark \ in $r$ band. The field of view is $8 \times 8$ arcmin. Stars 1-5 represent the photometric sequence of \citet{xilouris2006} recalibrated by us (see Table~\ref{tab:cal}).}
    \label{fig:fc}
\end{figure}

Data were processed with standard reduction procedures. The magnitude of 3C~371 was obtained with aperture photometry and calibrated with respect to reference stars in the same field of the source. A photometric sequence including five stars was published by \citet{xilouris2006}, but only in the $B$ and $I$ bands. We then used data acquired at the Teide Observatory in good nights to derive new calibrations in all $BVRI$ bands for the same stars (see Fig.~\ref{fig:fc}). The standard magnitudes obtained by our procedure are reported in Table~\ref{tab:cal}. Values in $B$ and $I$ bands are in good agreement with those published by \citet{xilouris2006}.
Magnitude offsets between the various datasets, which were mainly due to different choices of the aperture radius for the source photometry, leading to the inclusion of different contributions from the host galaxy light,  were corrected by aligning all datasets to the trend traced by those that adopted a 7.5$\arcsec$ aperture radius. Data points with very large errors and clear outliers were removed. Sequences of largely scattered data taken close in time by the same telescope were binned over a few minutes intervals.

\footnotetext{\url{http://ps1images.stsci.edu/cgi-bin/ps1cutouts}}

\begin{table}
\centering
\caption{Standard $BVRI$ magnitudes (and their uncertainties) of the reference stars in the field of view of 3C~371 (see Fig.~\ref{fig:fc}).}
\label{tab:cal}
\resizebox{\columnwidth}{!}{%
\begin{tabular}{lccccc}
\hline
Star & $B$ & $V$ & $R$ & $I$ \\ \hline
1 & 14.97 (0.03) & 14.25 (0.02) & 13.79 (0.03) & 13.41 (0.01) \\
2 & 15.16 (0.02) & 14.61 (0.03) & 14.25 (0.04) & 13.94 (0.01) \\
3 & 16.05 (0.05) & 15.49 (0.04) & 15.13 (0.04) & 14.81 (0.02) \\
4 & 16.70 (0.05) & 16.00 (0.05) & 15.54 (0.05) & 15.14 (0.02) \\
5 & 15.88 (0.02) & 15.15 (0.03) & 14.69 (0.03) & 14.28 (0.01) \\
\hline
\end{tabular}
}
\end{table}
%two stars, C1 and C2, were calibrated by the Perugia blazar monitoring program in $VRI$. Star C1 corresponds to Star 1 by \citet{xilouris2006}. We used C1 and C2 to calibrate the Xilouris et al. stars in $BVRI$ bands.

In addition to the data reduction specified above, we have also taken into account the effect of the Galactic extinction. We corrected the observed magnitudes using the Galactic extinction values $A_{B}=0.123$, $A_{V}=0.093$, $A_{R}=0.074$ and $A_{I}=0.051$, extracted from the NASA/IPAC Extragalactic Database\footnote{\url{https://ned.ipac.caltech.edu/}} (NED) and reported by \cite{schlafly2011}. Moreover, we also converted the magnitude measurements into flux units in order to construct the optical SEDs presented in Sect.~\ref{sec5}.  For this, we made use of the zero-mag fluxes provided by \cite{bessell1998}.

Finally, we have also taken into account the emission introduced by the host galaxy. 
The $R$-band magnitude of the host galaxy and the radius containing half of the light, $r_{e}$, were estimated to be $R_{host}=14.22 \pm 0.03$ and $r_{e}=8.6\arcsec \pm  0.1\arcsec$ by \cite{nilsson2003}. 
We used a De Vaucouleurs profile \citep{devaucouleurs1948} to model the brightness profile of the host galaxy in order to estimate its flux contribution within the 7.5\arcsec aperture. This was found to correspond to $\sim$46\% of the total light of the host galaxy.
We adopted the SWIRE template\footnote{\url{https://www.iasf-milano.inaf.it/~polletta/templates/swire\_templates.html}} \citep{polletta2007} for a 13 Gyr old Elliptical Galaxy to infer the host contributions within a 7.5\arcsec aperture radius in the other bands, using the $R$-band brightness as normalization.
These contributions are reported in Table~\ref{tab:host_galaxy_contributions} for the different filters, before and after correcting for Galactic extinction.

\begin{table}
\begin{center}
\caption{Contribution of the host galaxy to the total observed flux in the different optical bands.}
\label{tab:host_galaxy_contributions}
\resizebox{\columnwidth}{!}{%
\begin{tabular}{cccc}
\hline
\multirow{2}{*}{Band} &  \multirow{2}{*}{$\lambda_{eff}$ [\AA]} &  Observed contained   & Dereddened contained \\
                  &   & flux [mJy] & flux [mJy] \\ \hline
  $I$     &     7980          &     4.260 & 4.465       \\ \hline
  $R$     &     6410          &     2.901 & 3.106         \\ \hline
  $V$     &     5450          &     1.958 & 2.133         \\ \hline
  $B$     &     4380          &     0.815 & 0.912       \\ \hline
\end{tabular}
}
\end{center}
\end{table}

The WEBT $R$-band data were also used to convert the TESS counts flux to $R$-band magnitudes. This calibrated $R$-band light curve is shown in the third panel of Fig.~\ref{fig:webt_LCs} with the WEBT data from the same temporal period. The calibration was performed as $m_{TESS}=-2.5 \log{(\text{SAP\_FLUX})} + m_{0}$, where $m_{0}$ is an offset calculated by comparing each TESS observing sector with simultaneous $R$-band WEBT data, accordingly to previous studies by \cite{raiteri2021,raiteri2021b}. As can be observed in Fig.~\ref{fig:webt_LCs}, the calibration leads to an excellent agreement between the WEBT and TESS light curves. The differences in the variability amplitude observed between the different optical filters and the TESS data are related to the spectral response function and the spectral range covered by TESS\footnote{\url{https://heasarc.gsfc.nasa.gov/docs/tess/the-tess-space-telescope.html}}. This spans from 600~nm to 1000~nm, covering all the spectral range of the $I$ band, most of the $R$ band, but it has only a partial overlap with the $V$ band and almost no overlap in the $B$ band. We note however that the calibration was performed with respect to the optical $R$ band owing to the expected wider variations at this wavelength in comparison to the $I$ band and the fact that is the best sampled among the WEBT light curves \citep[see][]{raiteri2021,raiteri2021b}. 
%This $R$-band calibrated TESS light curve will be used for the time series analyses performed in this paper, unless specified otherwise.

\subsection{WEBT optical polarimetric observations}
\begin{figure}
    \includegraphics[width=\columnwidth]{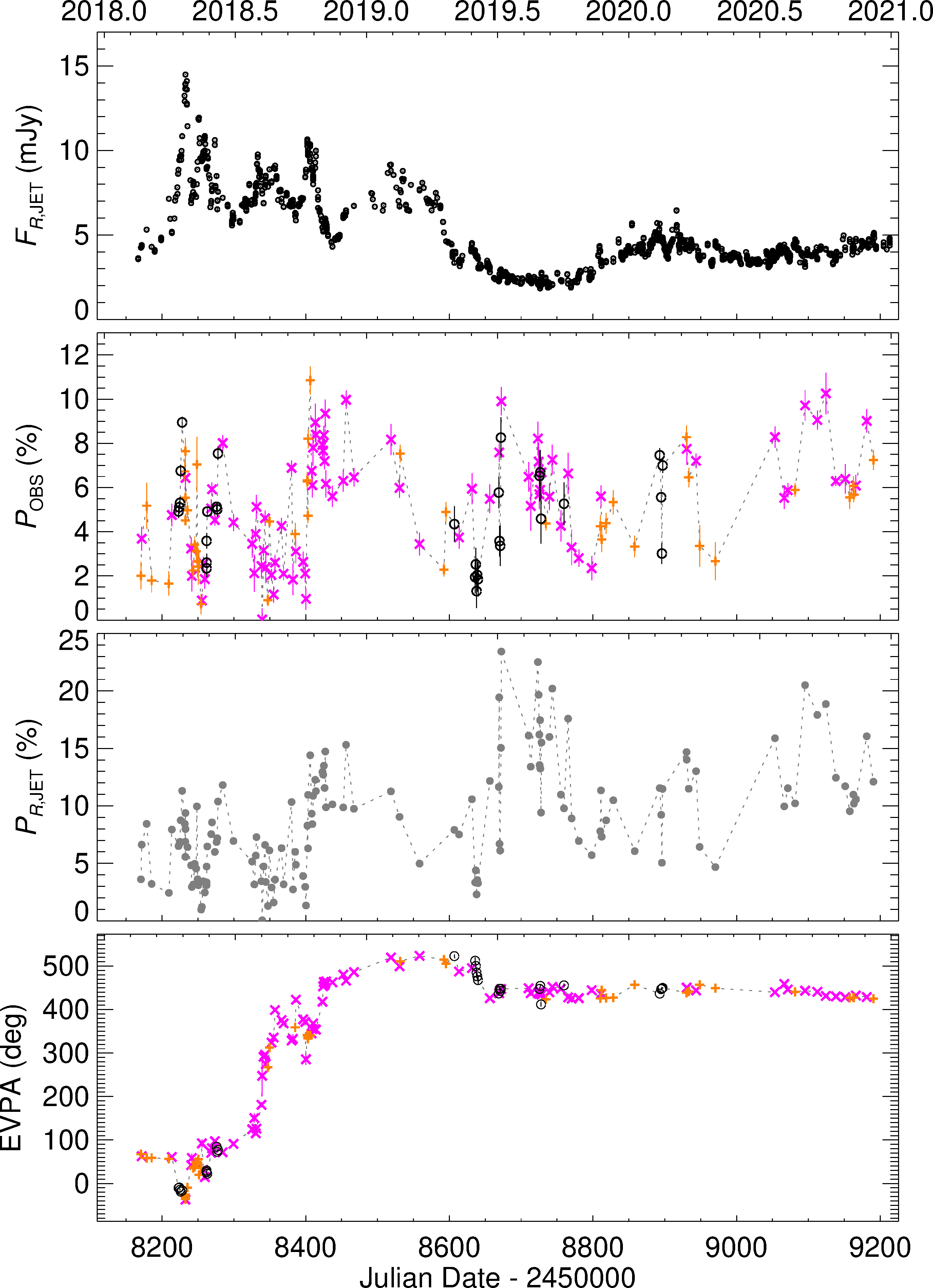}
    \caption{Optical polarization of 3C~371 during the $2018–2020$ period compared to the jet optical flux. \textit{From top to bottom:} $R$-band optical light curve (dereddened and host-galaxy-subtracted flux densities), observed polarization degree, intrinsic polarization degree, corrected EVPA behavior. Different colors and symbols are used to distinguish the contributing datasets, as specified in Table~\ref{tab:WEBT_optical_radio_obs}.}
    \label{fig:webt_polarization_LCs}
\end{figure}

In addition to the total optical observations in the $BVRI$ bands, the WEBT Collaboration has also performed a monitoring of the polarized emission of 3C~371 in the optical band making use of the telescopes listed in Table~\ref{tab:WEBT_optical_radio_obs} with available polarimetric instruments, that is, the Crimean, San Pedro Martir and St.~Petersburg Observatories. This monitoring obtained long-term measurements of the polarization degree and the electric vector position angle (EVPA) during the 3 year monitoring period. The EVPA has been corrected from the $\pm n \cdot 180^{\circ}$ ambiguity following the procedure typically used in the literature \citep[adding/subtracting $n \cdot 180^{\circ}$ to minimize the difference between consecutive measurements when $\Delta \theta_{i}= | \theta_{i+1}-\theta_{i} | >90^{\circ}$, see for instance][]{raiteri2023a,otero-santos2023b}. The observed polarization degree and EVPA trends in time are represented in Fig.~\ref{fig:webt_polarization_LCs} with the $R$-band light curve for comparison.

%We note that the polarization degree represented in Fig.~\ref{fig:webt_polarization_LCs} refers to the observed polarization degree. 
However, the contribution of the host galaxy to the total emission from blazars has a depolarizing effect of the polarization degree intrinsic to the jet \citep[see e.g.,][]{sosa2017}. Making use of the host galaxy contribution estimation performed in Sect.~\ref{sec2.2}, we calculated the intrinsic polarization degree as
\begin{equation}
P_{\text{jet}} \ [\%]=\frac{P_{\text{obs}} \ [\%]}{1-\frac{F_{\text{host}}}{F_{\text{total}}}},
\label{eq:depolarizing_effect}
\end{equation}
where $P_{\text{jet}}$ and $P_{\text{obs}}$ correspond to the intrinsic and observed polarization fractions, $F_{\text{total}}$ is the total brightness of the source and $F_{\text{host}}$ is the contribution of the host galaxy. This intrinsic polarization degree is also represented in Fig.~\ref{fig:webt_polarization_LCs}. We note that {no} bias correction was applied here to the polarization to account for the difference between the Rice and Gaussian distributions, as discussed by \cite{blinov2021}. Further discussion and analysis {of} the behavior of the polarization degree and EVPA of 3C~371 during this period is presented in Sect.~\ref{sec_polarization}.

\section{Intra-day variability}\label{sec3}
Several blazars in the past have shown signatures of IDV \citep[see e.g.,][]{wagner1995,villata2002,raiteri2008,gupta2008,gaur2015a,raiteri2021,raiteri2021b}. In particular for 3C~371, \cite{xilouris2006} have investigated the presence of IDV features in this source. Here, we take advantage of the extended temporal coverage provided by the TESS satellite for 3C~371 with a 2 min cadence to evaluate the possible IDV present in this BL Lac object. For this analysis we use the $R$-band calibrated light curve after Galactic extinction correction and host galaxy subtraction.

\subsection{Variability amplitude and timescales}
\begin{table*}
\centering
\caption{Results of the IDV analysis performed on each TESS sector during which 3C~371 was observed. These numbers correspond to the $R$-band scaled, dereddened and host-galaxy-subtracted magnitudes, and flux densities.}
\label{tab:idv_tess}
\begin{tabular}{ccccccccccc}
\hline
  TESS sector & $R_{max}$ & $R_{min}$ & $\Delta R_{max}$ & $F_{min}$ [mJy] & $F_{max}$ [mJy] & $\Delta F_{max}$ [mJy] & $A_{mp}$ [\%] & $\Delta A_{mp}$ [\%] \\ \hline
Sector 14  & 15.45 & 15.18 & 0.27 & 1.96 & 2.53 & 0.57 & 24.9 & 2.0  \\ \hline
Sector 15  & 15.51 & 15.14 & 0.37 & 1.86 & 2.61 & 0.75 & 34.1 & 2.1  \\ \hline
Sector 16  & 15.46 & 14.98 & 0.48 & 1.95 & 3.05 & 1.10 & 47.6 & 2.0  \\ \hline
Sector 17  & 15.46 & 14.97 & 0.49 & 1.95 & 3.05 & 1.10 & 44.9 & 1.3  \\ \hline
Sector 18  & 15.28 & 14.75 & 0.53 & 2.30 & 3.74 & 1.43 & 49.8 & 1.0  \\ \hline
Sector 20  & 14.84 & 14.48 & 0.35 & 4.80 & 3.47 & 1.33 & 32.6 & 0.6  \\ \hline
Sector 21  & 14.82 & 14.33 & 0.49 & 3.52 & 5.55 & 2.03 & 45.1 & 0.6  \\ \hline
Sector 22  & 14.75 & 14.24 & 0.51 & 3.77 & 6.02 & 2.25 & 49.6 & 0.8  \\ \hline
Sector 23  & 14.84 & 14.51 & 0.33 & 3.44 & 4.70 & 1.26 & 30.2 & 1.0  \\ \hline
Sector 24  & 14.93 & 14.58 & 0.35 & 3.17 & 4.40 & 1.23 & 32.9 & 0.7  \\ \hline
Sector 25  & 14.94 & 14.63 & 0.31 & 3.14 & 4.18 & 1.04 & 28.3 & 0.7  \\ \hline
Sector 26  & 14.95 & 14.68 & 0.27 & 3.12 & 3.99 & 0.87 & 24.8 & 0.9  \\ \hline
All data & 15.51 & 14.24 & 1.27 &  1.86 & 6.02 & 4.16 & 123.4 & 1.1 \\ \hline
\end{tabular}
\end{table*}

\begin{figure}
\centering
\includegraphics[width=\columnwidth]{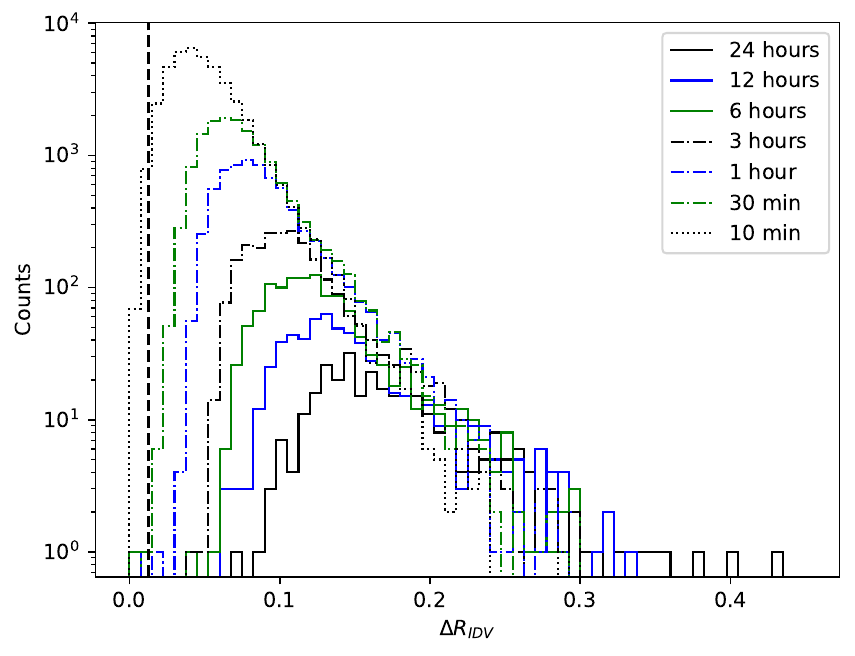}
\caption{Distributions of the maximum IDV magnitude variations observed from TESS data in different time intervals, which are distinguished with  different colors and line styles. The vertical black dashed line represents the mean uncertainty of the data.}
\label{fig:maximum_variability_histograms}
\end{figure}

Several episodes of noticeable IDV in the $R$-band emission of 3C~371 are observed over the period monitored by TESS in the R band emission. Some remarkable periods of IDV can be observed during different flares occurring on MJD~58741, MJD~58803 or MJD~58916 as displayed in Fig.~\ref{fig:tess_flux}. The total variability amplitude over the roughly 1 year period is $\Delta R_{\text{max}}=1.27$ mag, varying between magnitudes $R_{\text{max}}=15.51$ and $R_{\text{min}}=14.24$. We also evaluated the maximum magnitude intraday variations within different time intervals ranging from 24 hours down to 10 minutes. 
In Fig.~\ref{fig:maximum_variability_histograms} we represent the distributions of the maximum magnitude variations for the different time intervals considered here.
The peak of the distribution shifts towards higher values of $\Delta R_{\text{max}}^{\text{IDV}}$ for longer time intervals.
%, meaning that the amplitude of the fastest variations displayed by 3C~371 is smaller. 
The maximum noticeable variation that we observe in our entire dataset within a time interval of 10 minutes is $\Delta R^{\text{IDV}}_{\text{max}}=0.204$~mag, while changes of $\Delta R^{\text{IDV}}_{\text{max}}=0.431$~mag are observed within a 24-hour interval. This is also visible in the tails of the distributions shown in Fig.~\ref{fig:maximum_variability_histograms}, with for example the 6-hour histogram reaching values of $\Delta R^{\text{IDV}}_{\text{max}} \sim 0.35$~mag, while the 10~minute distribution extends only up to $\Delta R^{\text{IDV}}_{\text{max}} \sim 0.25$~mag. We note that we have taken into account the gaps existent in the light curves due to the end of each observing sector, as well as the gaps introduced in the middle of each sector by the data transfer, {by not considering the magnitude difference between the last point of each sector and the first of the next one, and the last point of the first half and first of the second half of each sector}. 

We can quantify the amplitude of these intraday flux variations relative to the average emission by using the variability amplitude parameter, $A_{mp}(\%)$, following the definition from \cite{romero1999},
\begin{equation}
A_{mp}(\%)=\frac{100}{<F>}\sqrt{(F_{\text{max}}-F_{\text{min}})^{2}-2\sigma^{2}},
\label{eq:var_amp}
\end{equation}
where $<F>$ is the mean value of the flux in the considered interval, $F_{\text{max}}$ and $F_{\text{min}}$ are the maximum and minimum flux values measured, and $\sigma$ is the flux uncertainty. The associated uncertainty of $A_{mp}$ can be estimated as
\begin{equation}
\begin{split}
\Delta A_{mp}(\%)=100 \times \left(\frac{F_{\text{max}}-F_{\text{min}}}{<F>A_{mp}}\right) \times \\ 
\times \sqrt{\left( \frac{\sigma_{\text{max}}}{<F>} \right)^{2} + \left( \frac{\sigma_{\text{min}}}{<F>} \right)^{2} + \left( \frac{\sigma_{<F>}}{F_{\text{max}}-F_{\text{min}}} \right)^{2}A_{mp}^{4}}.
\end{split}
\label{eq:var_amp_err}
\end{equation}
In this last equation, $\sigma_{<F>}$ corresponds to the standard deviation of the mean flux value. Considering the complete 2-minute cadence TESS light curve, the estimated value for the variability amplitude is $A_{mp}=(123.4 \pm 1.1)$\%. We report in Table \ref{tab:idv_tess} the derived values of $A_{mp}$ for the entire dataset, as well as each sector individually. We observe than on shorter timescales of $\sim$1 month, i.e. the duration of each sector, the variability is smaller, with variability amplitudes between $\sim$20\% and $\sim$50\% for the least and most variable sectors, respectively.
%\cite{pininti2023} report a fractional variability estimation for all TESS sectors of 3C~371, estimated though with the electron flux. Therefore, even if the values are not directly comparable to our estimations, we find the same pattern of most/least variable sectors. 

Moreover, we also estimated $A_{mp}$ for the flux density variations occurring in time intervals of 10~minutes, 30~minutes, 1~hour, 3~hours, 6~hours, 12~hours and 1~day. The results are displayed in Table~\ref{tab:amp_time_intervals_and_ANOVA} for each time interval considered. We observe that for the time intervals evaluated here, the amplitude of the variations is $\lesssim$13.08\%, with maximum changes between $\sim$19\% and $\sim$36\%, depending on the width of the interval. \cite{xilouris2006} report a very low amplitude of the variations ($\sim$1--1.5\%) on timescales as short as a few hours. {While this value is lower with respect to our estimations, there are some differences in the analyses that are probably leading to this discrepancy. \cite{xilouris2006} perform a 1.5-hour binning that may be smoothing real variations. Their observing cadence is also much lower than that from TESS data. Finally, the flux contribution of the host galaxy, not subtracted in the study by \cite{xilouris2006} can also be leading to a lower estimation of the variability of the jet}.
%, in line with the relatively small variability observed on short timescales here. 

\begin{table*}
\centering
\caption{Results of the variability amplitude parameter analysis and the ANOVA test performed within different time intervals.}
\label{tab:amp_time_intervals_and_ANOVA}
\begin{tabular}{cccccccc}
\hline
$\Delta t$ & 10 minutes & 30 minutes & 1 hour & 3 hours & 6 hours & 12 hours & 24 hours \\ \hline
$N_{\text{intervals}}$ & 43776 & 14592 & 7296 & 3648 & 1216 & 608 & 304 \\ \hline
$<A_{mp}>$ [\%] & 3.25 $\pm$ 0.01 & 4.79 $\pm$ 0.02 & 6.05 $\pm$ 0.03 & 7.92 $\pm$ 0.05 & 9.22 $\pm$ 0.08 & 10.82 $\pm$ 0.13 & 13.08 $\pm$ 0.24 \\ \hline
$A_{mp}^{max}$ [\%] & 19.3 $\pm$ 3.8 & 20.8 $\pm$ 2.8 & 20.7 $\pm$ 2.7 & 20.5 $\pm$ 2.7 & 23.5 $\pm$ 2.5 & 31.2 $\pm$ 1.7 & 36.0 $\pm$ 0.9 \\ \hline
$N_{\text{intervals}}^{\text{p-value}<0.001}$ & -- & 11 (0.07\%) & 24 (0.33\%) & 223 (6\%) & 442 (36\%) & 426 (70\%) & 287 (94\%) \\ \hline
\end{tabular}
\end{table*}

In order to quantify the fastest variability that we are able to significantly resolve, we follow a similar approach to that from \cite{weaver2020}, where an ANOVA (ANalysis Of VAariance) test was implemented to evaluate the IDV of BL~Lacertae. This test has also been used in other variability studies in blazars \citep[see e.g.,][]{dediego2010,fraija2017,weaver2019}. We divide the dataset in intervals of different width. Then, breaking each interval in subgroups, we evaluate the null hypothesis of every subgroup having the same mean value, which can be translated into anonvariable nature. When the contrary occurs with a confidence of $\sim$3$\sigma$ (p-value~$\leq$~0.001), we can assume that the source is variable on timescales shorter than the interval considered \citep[see for more details][]{weaver2020}. We report in Table~\ref{tab:amp_time_intervals_and_ANOVA} the number of intervals where we can reject the null hypothesis of non-variability for the different time interval widths considered.

\begin{figure}
        \includegraphics[width=\columnwidth]{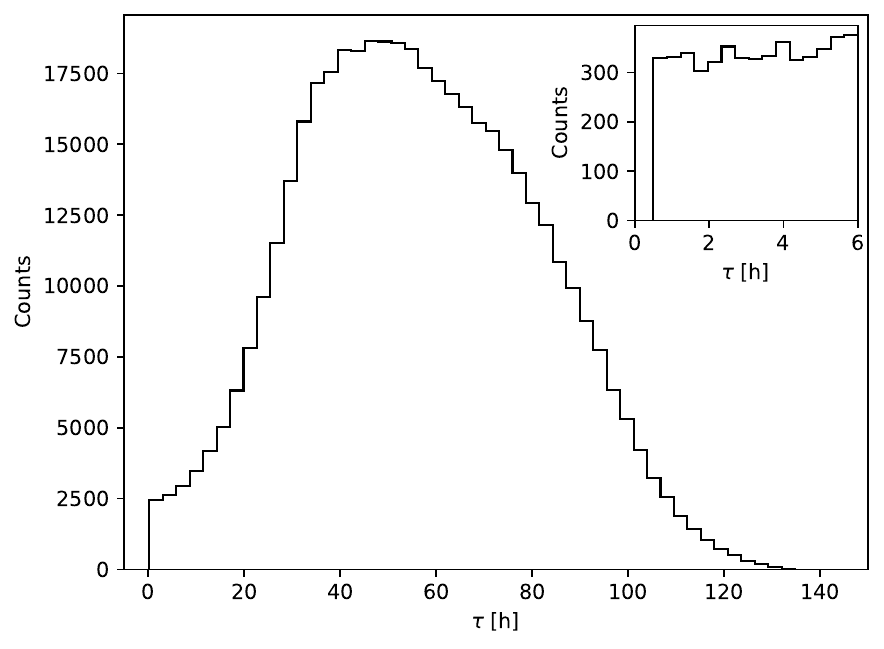}
    \caption{Histogram of the minimum variability timescales measured from the variable intervals according to the ANOVA test performed here. The inner panel shows a zoomed-in histogram for timescales of shorter than 6~hours.}
    \label{fig:minimum_variability_timescales}
\end{figure}

We observe that for most of the intervals corresponding to $\Delta t \leq 1$~hours, we barely see any sign of significant variability, with only 24 one-hour intervals fulfilling the aforementioned 3$\sigma$ variability criterion. On the other hand, for $\Delta t \geq 6$~hours we can reject the nonvariable null hypothesis, as at least $\sim$36\% of the intervals show significant variability according to the ANOVA test. Considering the 6-hour intervals, we can determine the shortest variability timescale that we are able to resolve in our dataset following the procedure from \cite{weaver2020} \citep[for more details see also][]{burbidge1974}. This procedure determines, for all pairs of points of the subset of variable 6-hour time intervals with $S_2 - S_1 > 3(\sigma_1 + \sigma_2)/2$ and $S_2 > S_1$, their characteristic variability timescale as
\begin{equation}
\tau = \Delta t / \ln{(S_2/S_1)},
\label{eq:tau_variability}
\end{equation}
where $S_1$ and $S_2$ are the flux values of each point of the pair, $\sigma_1$ and $\sigma_2$ are their corresponding uncertainties, and $\Delta t = |t_2 - t_1|$ is the time separation between them. 
We note that this method is applied on the flux measurements rather than on the magnitudes. Therefore, as in \cite{weaver2020}, we use for this the TESS dereddenned, host galaxy subtracted flux density light curve.
In Fig.~\ref{fig:minimum_variability_timescales} we represent a histogram of the characteristic variability timescales determined with Eq.~(\ref{eq:tau_variability}). While most of the variability shows timescales longer than 1 day, with a mean value of $< \tau > \ = 56.5$~hours and a standard deviation of 24.5~hours, we can still observe significant IDV signatures. This is expected, given the long-term coverage of the TESS data, spanning almost 1~year and therefore allowing us to observe variations on much longer timescales than IDV. Here we focus on the significant IDV signatures identified with the ANOVA analysis and timescale quantification commented above. {The fastest variability we were able to measure is quite remarkable}, with a characteristic timescale of $\tau_{\text{min}}=0.47$~hours.

Apart from this fast variability, and as can be observed in the TESS light curve represented in Fig.~\ref{fig:tess_flux}, several periods of variability are characterized with different variability timescales longer than the minimum variability timescale estimated above. For evaluating the presence of other timescales involved in the IDV observed, we used the Fourier-based technique of the wavelet analysis, which has long been used as a powerful tool for evaluating in the past variability timescales in blazars, owing to its ability to decompose the signal in the two-dimensional time-frequency space, allowing the user to find the different timescales present in a dataset, and contextualize them in time \citep[see][]{gupta2009,zhou2018,raiteri2023a}. When these timescales are persistent in time, they reveal the presence of a periodic variability of the data. Hence, this tool has also been widely used for performing periodicity searches in the emission of blazars {\citep[e.g.,][]{otero-santos2020,penil2020,jorstad2022,roy2022,otero-santos2023,li2023,tripathi2024}}. Moreover, it has been also employed for identifying nonperiodic characteristic variability timescales in blazar light curves \citep[see][]{raiteri2023a}. Here we use the \textsc{python} implementation of the wavelet function code developed by \cite{torrence1998}\footnote{\url{https://github.com/ct6502/wavelets}}.

\begin{figure}
        \includegraphics[width=\columnwidth]{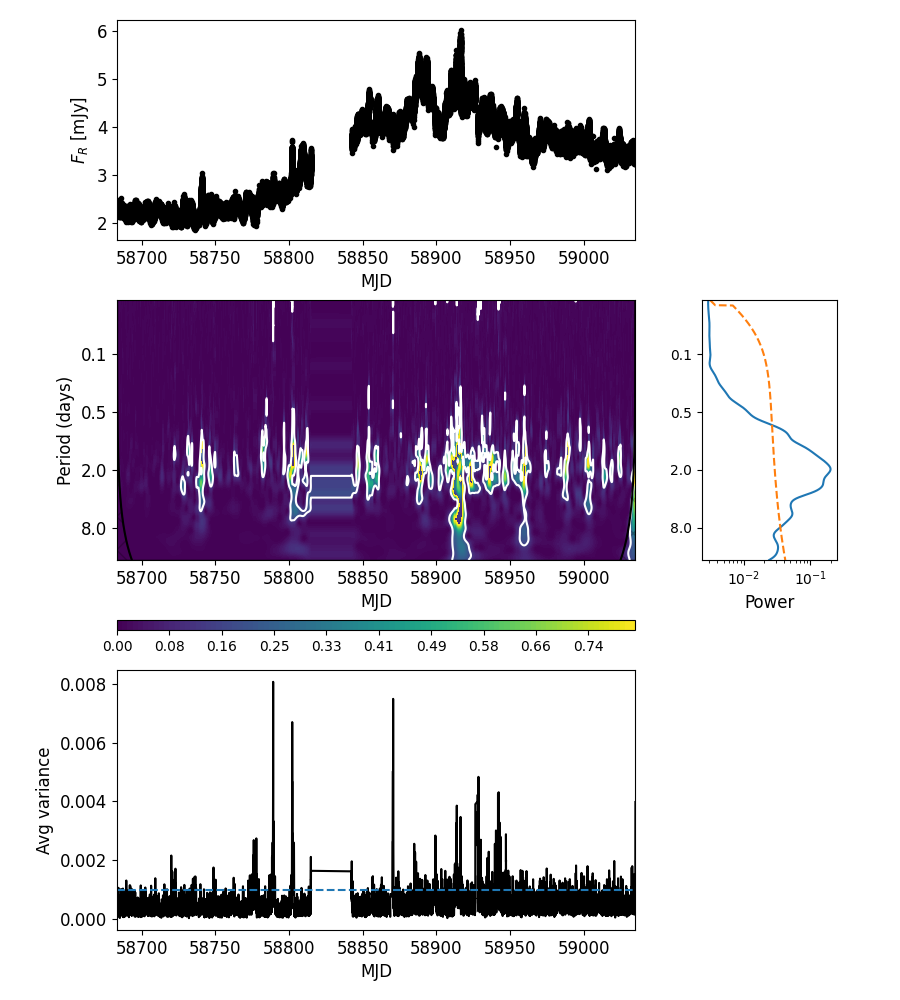}
    \caption{IDV wavelet analysis performed on the dereddened, host-galaxy-subtracted {$R$-band flux light curve} of 3C~371. \textit{Top}: $R$-band light curve. \textit{Middle}: Time-period power spectrum (left) and time-folded power spectrum (right). White contours in the left panel and the dashed line in the right panel represent variability signatures with a significance >$3\sigma$ (>99.73\% confidence level). \textit{Bottom:} Averaged variance between 4~minutes and 1~day. The horizontal dashed line represents the 3$\sigma$ (99.73\%) confidence level.}
    \label{fig:wavelet_IDV_alldata}
\end{figure}

It is well known that long-term trends can mask the impact of variability on shorter timescales \citep[see][]{raiteri2021,raiteri2021b}. Owing to the roughly 1 year time span of the TESS data, we have modelled and subtracted variability signatures on timescales of longer than 1~day, which are further studied using the 3 year data from WEBT (see Sect.~\ref{sec4}). Following the procedure employed by \cite{raiteri2021,raiteri2021b}, we model the long-term variations through a cubic spline interpolation of a binned light curve. This cubic spline interpolation has been widely used in blazar studies to represent the observed LTV in these sources \citep[e.g.,][]{ghisellini1997,raiteri2017}. In our case, we perform the interpolation on the daily binned TESS light curve. In Fig.~\ref{fig:wavelet_IDV_alldata} we show the derived wavelet spectrum with associated frequency-space power spectral density (PSD).

\begin{figure*}
\centering
    \includegraphics[width=0.49\textwidth]{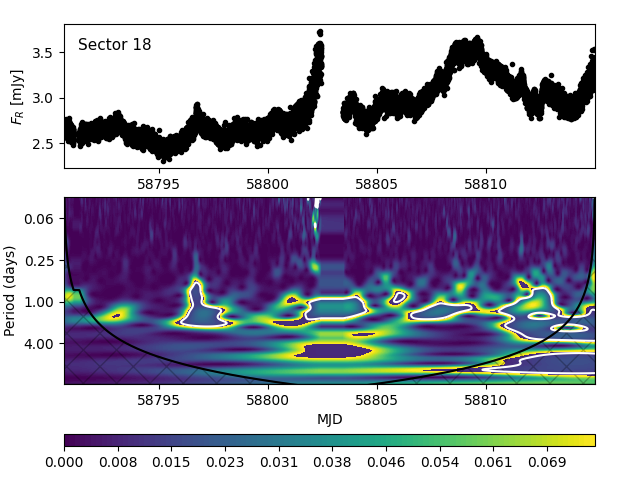}
    \includegraphics[width=0.49\textwidth]{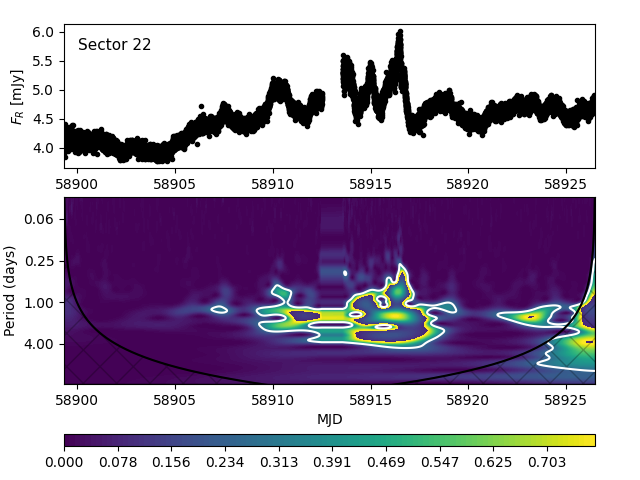}
    \caption{IDV wavelet analysis performed on the dereddened, host-galaxy-subtracted {$R$-band flux light curves} of 3C~371 for sectors 18 (\textit{left}) and 22 (\textit{right}). \textit{Top}: $R$-band light curves. \textit{Bottom}: Time-period power spectra. White contours represent variability signatures with a significance >$3\sigma$ (>99.73\% confidence level). Gray grids represent the cone of influence regions affected by edge effects.} 
    \label{fig:wavelet_IDV_sectors}
\end{figure*}

We observe both from the time-period spectrum (middle left panel) and the period-folded PSD (middle right panel) represented in Fig.~\ref{fig:wavelet_IDV_alldata} that a large amount of the IDV variability of 3C~371 appears on timescales between 0.5~days and 3-5~days. Nevertheless, several features of faster variations  are visible during different periods, most of them coincident with bright, fast developing flares (e.g. on MJD~58916, approximately). A wide variety of intraday characteristic timescales are easily identified at a 3$\sigma$ confidence level. As an example, we represent the wavelet analysis performed on sectors 18 and 22 individually, where interesting flaring events on very fast timescales can be identified. The results of these analyses are shown in Fig.~\ref{fig:wavelet_IDV_sectors}. Again, we see several signatures on timescales ranging from 0.5~days to $\sim$3~day. In addition, we are able to resolve even faster variations during bright flares occurring in these particular sectors. For instance, during the flare on MJD~58916, the power spectrum reveals significant flux variability down to timescales <0.25~days (see right panels in Fig.~\ref{fig:wavelet_IDV_sectors}). Another example is the flare on MJD~58802 (left panels in Fig.~\ref{fig:wavelet_IDV_sectors}), where we can observe one of the fastest signatures of 3C~371 with this analysis, showing a characteristic timescale $\leq$0.05~days, compatible with the fastest signatures found with the ANOVA approach.

Moreover, also in line with the ANOVA results, we see significant flux variations on timescales longer than 1~day. However, as mentioned above, we go into more detail on longer timescales in Sect.~\ref{sec4}. Hence, the period scale of the IDV wavelet analysis is truncated at 8~days. We also note that, due to the gap present in the complete TESS light curve during sector 19, when the source was not in the field of view, an artificial signature of variability appears in the wavelet representations, which should be ignored as it does not represent a real characteristic variability timescale.

Minute and hour timescale variability has been attributed in the past to changes related to turbulence of plasma in the emitting jet region. This fast variability suggests a very compact emitting region. Therefore, the minimum variability timescale and the causality condition can be used to set a constraint on the size of the region responsible for such emission and variability, given by \citep[see e.g.,][]{aharonian2007}
\begin{equation}
R\leq \frac{\tau_{min} c \delta }{1+z},
\label{eq:radius}
\end{equation}
where $\tau_{min}$ is the variability timescale used for this estimation, $c$ is the speed of light, $\delta$ represents the Doppler factor of the emitting region and $z$ is the redshift of the source. We use the Doppler factor $\delta=9.5$ reported by \cite{chen2018} and the value of $\tau_{\text{min}}=0.47$~hours estimated here. Under these assumptions, the estimated size of the emitting region responsible for the IDV observed from  3C~371 is $R < 6.5 \times 10^{14}$~cm (i.e. $\sim$$2\times 10^{-3}$~pc), roughly a factor 2 more constraining than the estimation performed by \cite{carini1998}. This size is smaller than the general emitting region size typically assumed for blazars, as discussed by \cite{raiteri2023b} for the case of S4~0954+65. Therefore, the observed variability can likely be ascribed to flux fluctuations within a subregion of the relativistic jet.

Moreover, the minimum variability timescale can also be used to make a rough estimation of the mass of the central black hole as proposed in previous works \citep{xie2002b,xie2002,dai2007,gupta2012}
\begin{equation}
M_{BH}\leq 1.62 \times 10^{4} \frac{\tau_{min} \delta }{1+z} M_{\odot},
\label{eq:BH_mass}
\end{equation}
where $M_{BH}$ and $M_{\odot}$ are the masses of the black hole and the Sun, respectively. Using the aforementioned Doppler factor and $\tau_{\text{min}}$ values, we derive an estimate of the black hole mass $M_{BH} \leq 3.4 \times 10^{8}M_{\odot}$, which is consistent with the typical mass ranges of $10^{6} - 10^{10} M_{\odot}$ assumed for supermassive black holes in AGN \citep[see for instance][]{kormendy1995,richstone1998}, and >$10^{8} M_{\odot}$ for the case of radio-loud AGN \citep{chiaberge2011}. This value is also consistent within errors with different estimations reported by other authors \citep[see e.g.,][]{dai2007,chen2018,pei2022}.

\subsection{Power spectral density analysis}
\begin{figure*}
\centering
    \subfigure{\includegraphics[width=0.67\columnwidth]{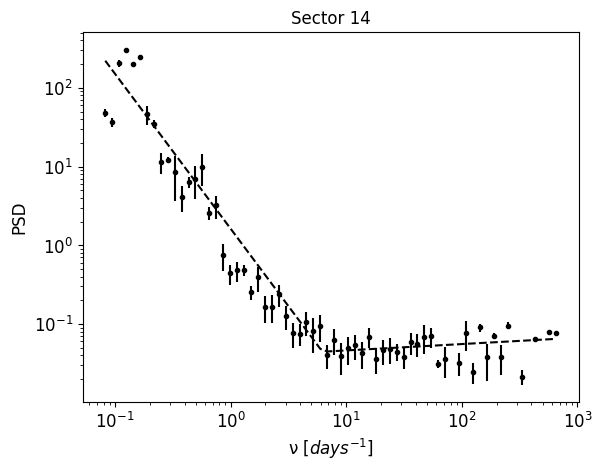}}
    \subfigure{\includegraphics[width=0.67\columnwidth]{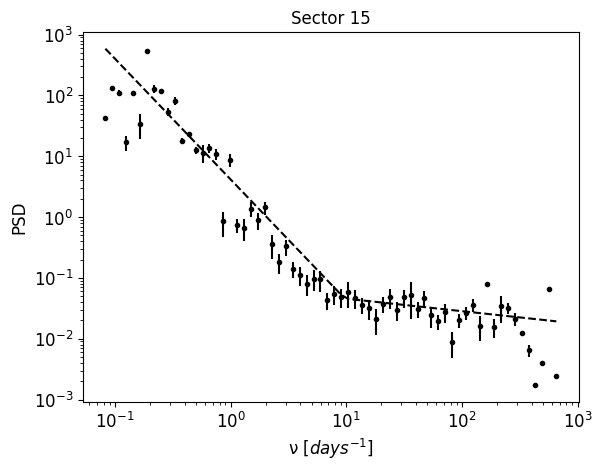}}
    \subfigure{\includegraphics[width=0.67\columnwidth]{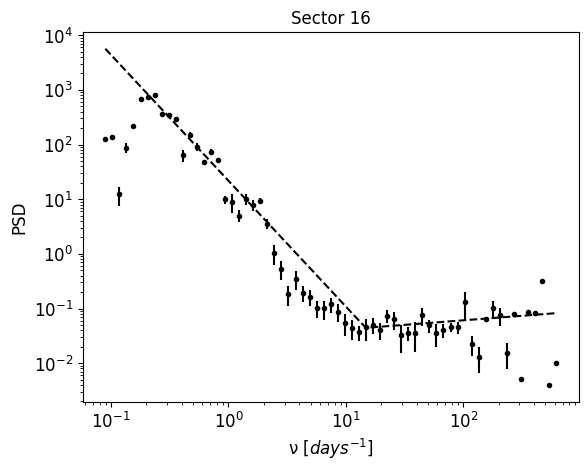}}
    \subfigure{\includegraphics[width=0.67\columnwidth]{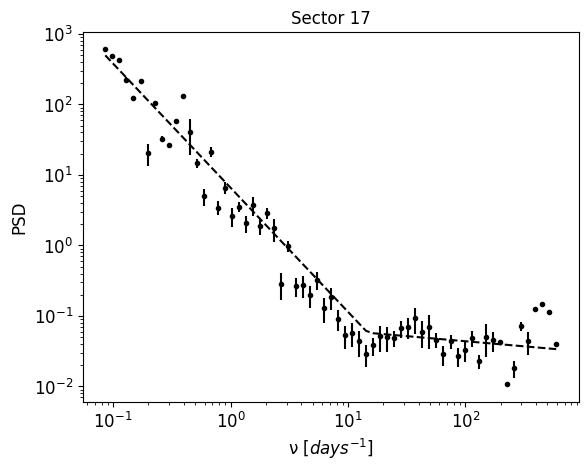}}
    \subfigure{\includegraphics[width=0.67\columnwidth]{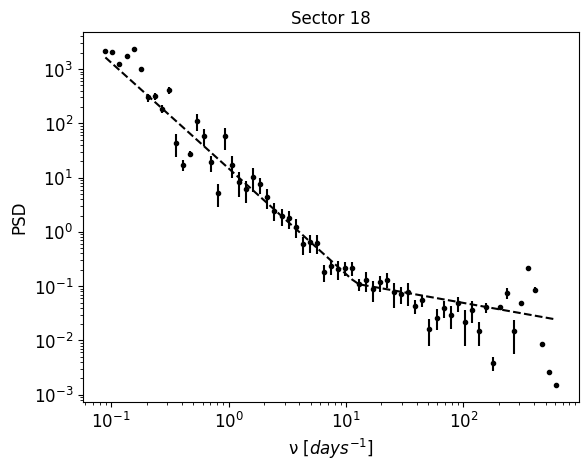}}
    \subfigure{\includegraphics[width=0.67\columnwidth]{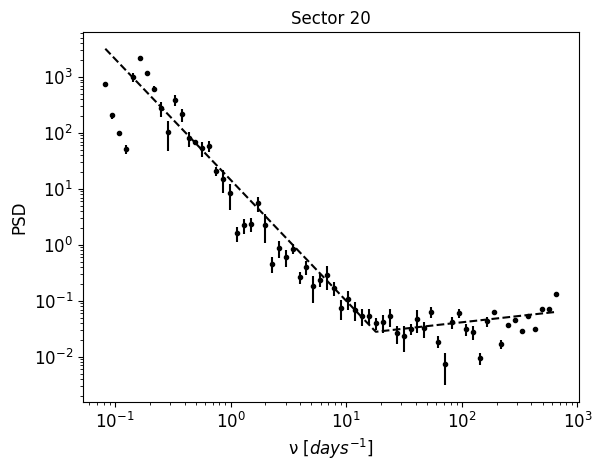}}
    \subfigure{\includegraphics[width=0.67\columnwidth]{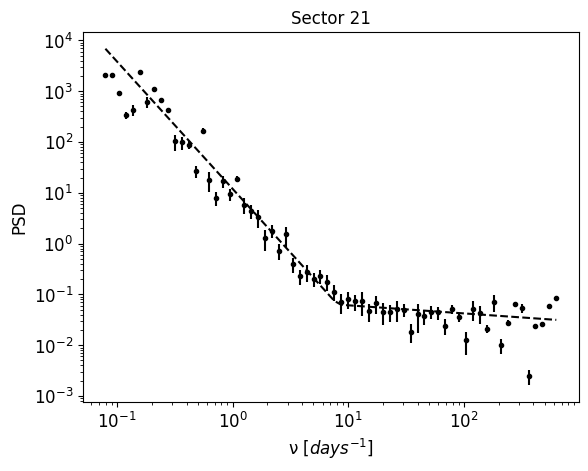}}
    \subfigure{\includegraphics[width=0.67\columnwidth]{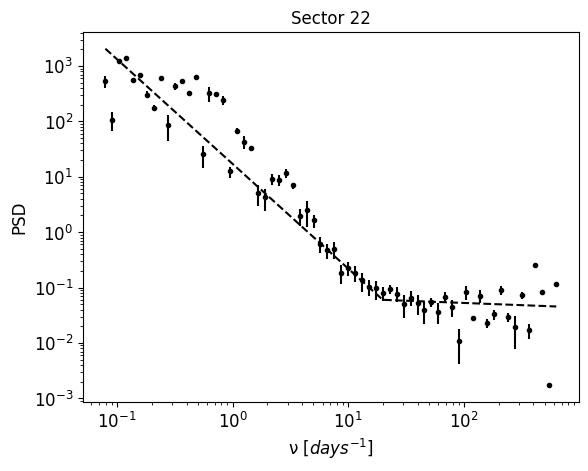}}
    \subfigure{\includegraphics[width=0.67\columnwidth]{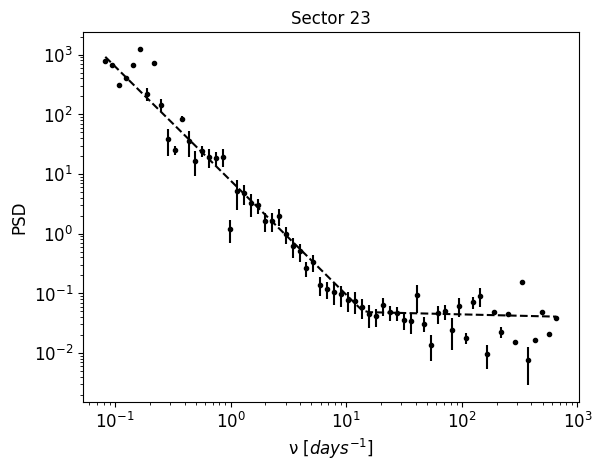}}
    \subfigure{\includegraphics[width=0.67\columnwidth]{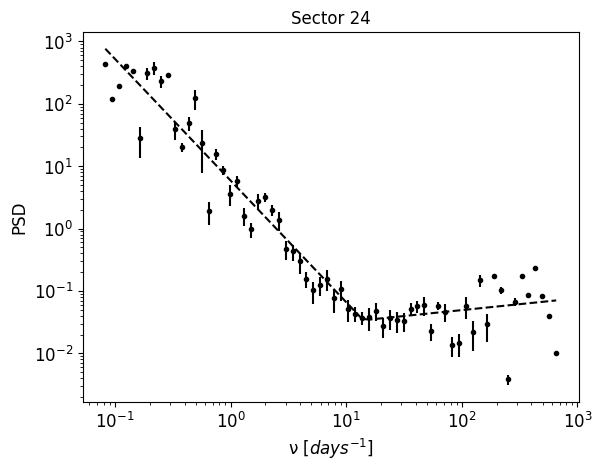}}
    \subfigure{\includegraphics[width=0.67\columnwidth]{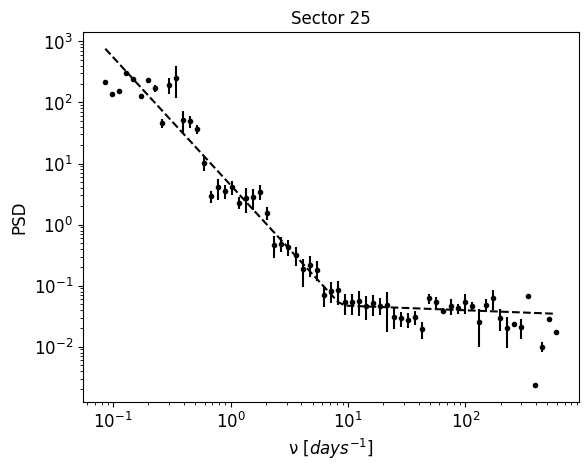}}
    \subfigure{\includegraphics[width=0.67\columnwidth]{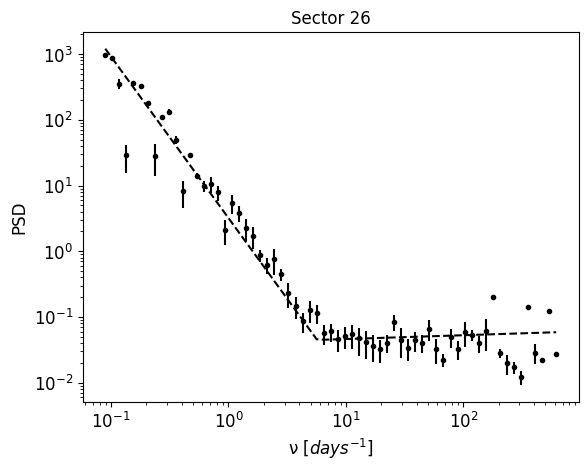}}
    \caption{Binned PSDs of the different TESS sectors. The dashed black lines represent the broken power law spectral shapes used for modeling the PSDs reported in Table \ref{tab:psd_tess}.} 
    \label{fig:PSD_analysis_TESS}
\end{figure*}

We conducted an evaluation of the PSD associated to the data taken by TESS. PSD analyses in blazars have turned to be extended and effective tools for evaluating the type and origin of the variability, characteristic timescales, connection of the variability with the accretion disk, as well as the nature of the underlying noise \citep[see for instance][]{bhatta2020,goyal2022}. For these reasons we calculated the PSD corresponding to each TESS sector. Due to the frequency sampling, the PSD is better sampled at low frequencies with respect to high frequencies. To account for this difference, we binned the derived PSDs with an equispaced binning of 0.06~Hz in logarithmic space, leading to a regularly sampled PSD. All the PSDs are represented in the different panels of Fig.~\ref{fig:PSD_analysis_TESS}. 

A common way of modeling the PSD of blazars has been through power law functions \citep[$\mathcal{P} \propto \nu^{-\alpha}$, see for instance][]{bhatta2020,goyal2022}, where the value of the spectral index provides information of the type of noise dominating the variability of the data ($\alpha=-2$ corresponding to red noise, $\alpha=-1$ representing flicker noise and $\alpha=0$ indicating white noise). In fact, typical values of blazar PSDs tend to be close to $\alpha=-2$, indicating a dominant red noise on long timescales. Moreover, when the cadence of the observations allow us to study the shortest timescales, a drastic flattening of the slope is often observed. In these cases, a broken power law function is typically used to describe the PSD, accounting for this spectral index change at high frequencies.

Under these considerations, owing to the regular and high cadence of the TESS data, we modeled the PSD of our data using a broken power law function defined as
\begin{equation}
\mathcal{P} = \left\{ \begin{array}{lc} A(\nu/\nu_{b})^{-\alpha_{1}}: & \nu<\nu_{b} \\ \\  A(\nu/\nu_{b})^{-\alpha_{2}}: & \nu>\nu_{b}  \end{array} \right.
\label{eq:broken_power_law}
\end{equation}
where $\alpha_1$ and $\alpha_2$ are the spectral indices below and above the break frequency $\nu_{b}$, and $A$ is the amplitude at $\nu_{b}$. The need for this spectral shape is also clear from visual inspection of the PSDs represented in Fig.~\ref{fig:PSD_analysis_TESS}, where a change of the spectral index is clearly observable at high frequencies. The best-fit parameters derived are reported in Table~\ref{tab:psd_tess}. 

All the PSDs derived from the TESS data are well represented by a broken power law function, where the low-frequency power can be modeled with a red-noise power law ($\alpha \sim 2$). At frequencies between 5.69~days$^{-1}$ and 19.88~days$^{-1}$ (timescales between $\sim$4.2~hours and $\sim$1.2~hours) we observe a break between the low- and high-frequency regime, where white noise becomes dominant, as observed from the spectral index $\alpha \sim 0$ above the break. This flattening and white-noise dominance has been observed in the past for IDV PSDs of blazars on the shortest variability timescales \citep[e.g.,][]{raiteri2021,raiteri2021b,gowtami2022}, and has been interpreted as a characteristic variability timescale, which is found to be within the range of fast variation signatures reported in the previous section. Therefore, IDV and STV at frequencies below the break can be related to a red-noise like, stochastic nature of the synchrotron variability, most likely due to physical processes in the jet itself within small emitting regions, supporting the scenario where these variations come from a jet subregion \citep{ryan2019}. Above $\nu_{b}$, white noise overcomes the power variability of red noise. These breaks could be related with electron escape, cooling, or light-crossing timescales \citep{ryan2019}.

\begin{table}
\centering
\caption{Results of the PSD analysis performed on each TESS sector during which 3C~371 was observed.}
\label{tab:psd_tess}
\resizebox{\columnwidth}{!}{%
\begin{tabular}{ccccc}
\hline
TESS sector & $A$ & $\alpha_{1}$ & $\alpha_{2}$ & $\nu_{b}$ [days$^{-1}$]   \\ \hline
Sector 14  & $0.044 \pm 0.036$ & $1.97 \pm 0.27$ & $-0.08 \pm 0.21$ &  $6.23 \pm 1.84$ \\ \hline
Sector 15  & $0.046 \pm 0.019$ & $1.98 \pm 0.17$ & $0.20 \pm 0.41$ &  $9.68 \pm 0.20$   \\ \hline
Sector 16  & $0.041 \pm 0.017$ & $2.30 \pm 0.21$ & $-0.20 \pm 0.38$ &  $15.28 \pm 0.29$ \\ \hline
Sector 17  & $0.058 \pm 0.012$ & $1.76 \pm 0.34$ & $0.14 \pm 0.36$ &  $14.74 \pm 0.79$ \\ \hline
Sector 18  & $0.110 \pm 0.058$ & $1.95 \pm 0.09$ & $0.39 \pm 0.29$ &  $12.38 \pm 3.51$ \\ \hline
Sector 20  & $0.028 \pm 0.012$ & $2.16 \pm 0.14$ & $-0.23 \pm 0.18$ & $18.02 \pm 3.31$  \\ \hline
Sector 21  & $0.063 \pm 0.034$ & $2.51 \pm 0.32$ & $0.16 \pm 0.45$ &  $8.11 \pm 3.53$ \\ \hline
Sector 22  & $0.060 \pm 0.013$ & $1.89 \pm 0.18$ & $0.08 \pm 0.11$ & $19.88 \pm 2.93$  \\ \hline
Sector 23  &  $0.048 \pm 0.040$ & $1.91 \pm 0.08$ & $0.05 \pm 0.37$ &  $14.19 \pm 4.20$ \\ \hline
Sector 24  & $0.034 \pm 0.022$ & $1.95 \pm 0.35$ & $-0.19 \pm 0.39$ & $14.06 \pm 2.48$  \\ \hline
Sector 25  & $0.047 \pm 0.017$ & $2.09 \pm 0.19$ & $0.07 \pm 0.39$ &  $8.76 \pm 1.69$ \\ \hline
Sector 26  & $0.045 \pm 0.021$  & $2.45 \pm 0.30$ & $-0.05 \pm 0.16$ & $5.69 \pm 1.57$  \\ \hline
\end{tabular}
}
\end{table}

\section{Long-term light curves}\label{sec4}
Apart from the remarkable very fast variability observed from 3C~371 thanks to the extraordinarily high observing cadence of TESS, this source also displays variations in its emission on longer timescales. Using the 3 year optical light curves provided by the WEBT Collaboration presented in Sect.~\ref{sec2}, we study the optical LTV observed from this blazar.

\subsection{Variability}
During the 3 year monitoring of 3C~371 we can observe different emission states. The first year, the variability is mainly dominated by several flares (see Fig.~\ref{fig:webt_LCs}). After this period, on MJD~58500 approximately, the source experienced a significant decrease of its emission, leading to a low and much less variable emission state. After this period, a relatively steady increasing trend in the flux was detected. Finally, a state of fairly constant emission on long timescales was reached after MJD~58900. 

As for the IDV, we characterized the amplitude of the observed LTV for the different optical bands. The results are reported in Table \ref{tab:ltv}. The maximum variation on long timescales of the $R$-band flux density is $\Delta F_{\text{max}}= 12.65$~mJy, with maximum and minimum values of $F_{\text{max}}=14.49$~mJy and $F_{\text{min}}=1.84$~mJy, and $A_{mp}= (264.4\pm13.9)$\%. All the optical bands report similar results, with a slightly lower variability in the $I$ band. 

\begin{table}
\begin{center}
\caption{Results of the LTV analysis performed on the optical emission of 3C~371. The data have been corrected for Galactic extinction and the host galaxy contribution has been subtracted.}
\label{tab:ltv}

\begin{tabular}{cccccc}
\hline
\multirow{2}{*}{Band} & $F_{min}$ & $F_{max}$ & $\Delta F_{max}$  & $A_{mp}$  & $\Delta A_{mp}$ \\
                  & [mJy] & [mJy] & [mJy] & [\%] & [\%] \\ \hline
$I$  & 2.38 & 16.15 & 13.78 & 203.0 & 43.5 \\ \hline
$R$  & 1.84 & 14.49 & 12.65 & 264.4 & 13.9 \\ \hline
$V$  & 1.16 & 10.09 & 8.92 & 276.3 & 21.9 \\ \hline
$B$  & 0.85 & 6.92 & 6.08 & 247.0 & 28.4 \\ \hline
\end{tabular}
\end{center}
\end{table}

We also calculated $A_{mp}$ within different time intervals following the same procedure as for the IDV and reported in Table~\ref{tab:amp_time_intervals_and_ANOVA}. The results for the LTV are presented in Table~\ref{tab:LTV_amplitude_intervals} for the optical $R$ band. As expected from the variability amplitude estimated in different IDV timescales and for the complete long-term light curves, $A_{mp}$ increases for longer time intervals, meaning that typically, LTV shows much larger amplitudes than IDV. 

\begin{table*}
\centering
\caption{Variability amplitude within different time intervals for the long-term optical $R$ band flux density light curve.}
\label{tab:LTV_amplitude_intervals}
\resizebox{\textwidth}{!}{%
\begin{tabular}{cccccccc}
\hline
$\Delta t$ & 10~days & 20~days & 30~days & 50~days & 100~days & 200~days & 300~days \\ \hline
$<A_{mp}>$ [\%] & $20.2 \pm 1.1$ & $31.7 \pm 1.9$ & $40.3 \pm 2.9$ & $54.3 \pm 4.2$ & $75.4 \pm 10.1$ & $112.8 \pm 19.6$ & $152.4 \pm 19.4$ \\ \hline
$A_{mp}^{max}$ [\%] & $48.9 \pm 18.2$ & $70.8 \pm 41.2$ & $78.6 \pm 15.7$ & $87.8 \pm 7.5$ & $124.9 \pm 36.3$ & $179.0 \pm 16.9$ & $197.66 \pm 23.7$ \\ \hline

\end{tabular}
}
\end{table*}

\begin{figure}
        \includegraphics[width=\columnwidth]{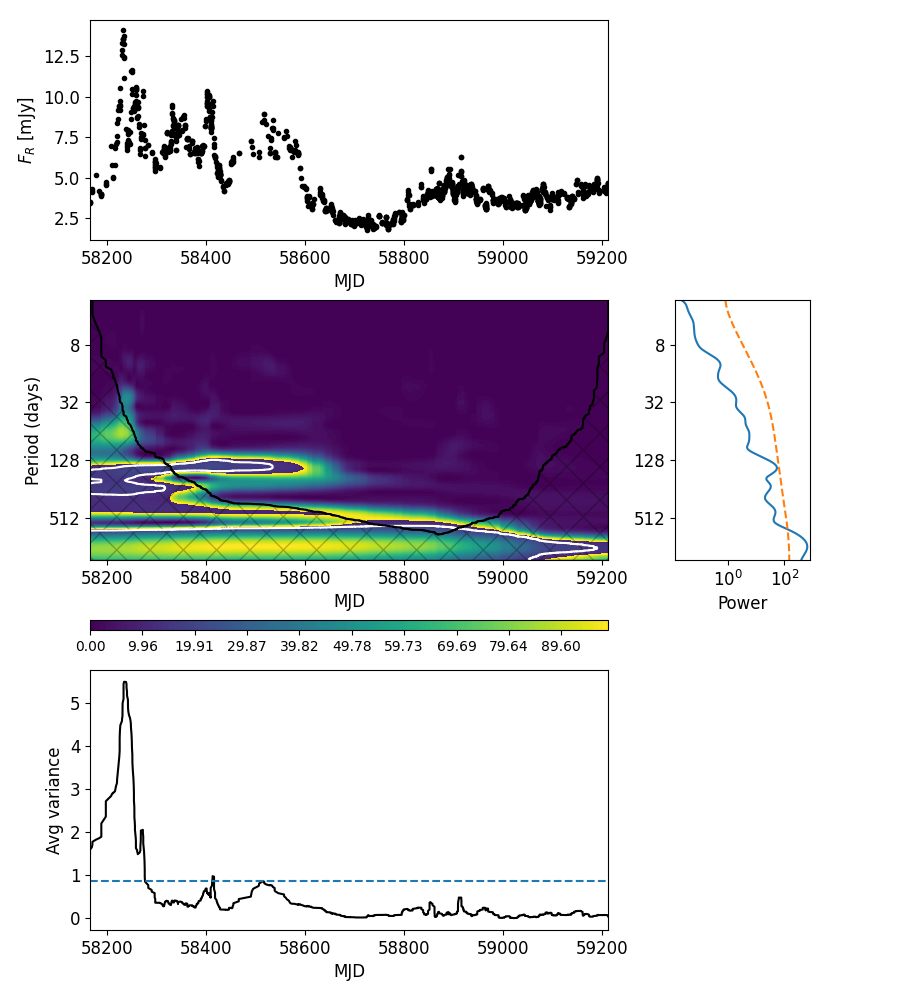}
    \caption{LTV wavelet analysis performed on the dereddened, host-galaxy-subtracted and flux-calibrated $R$-band light curve of 3C~371. 
    %Same description as Fig.~\ref{fig:wavelet_IDV_alldata}.
    \textit{Top}: $R$-band light curve. \textit{Middle}: Time-frequency power spectrum (left) and time-folded power spectrum (right). White contours in the left panel and the dashed line in the right panel represent variability signatures with a significance >$3\sigma$ (>99.73\% confidence level). The gray grid represents the cone of influence region affected by edge effects. \textit{Bottom:} Averaged variance between 1 and 100 days. The horizontal dashed line represents the 3$\sigma$ (99.73\%) confidence level.
    }
    \label{fig:wavelet_LTV_observed}
\end{figure}

The broadband emission of blazars often involves several physical processes with different characteristic variability timescales entangled. To evaluate these timescales, we used the wavelet method introduced above, which allows us to identify (and locate in time) periods of significant variation. We represent in Fig.~\ref{fig:wavelet_LTV_observed} the wavelet transform, with its associated PSD, corresponding to the $R$-band flux density light curve. 

However, as already introduced in Sect.~\ref{sec3}, blazar time series are typically dominated by the longest variability timescales present in the data, masking the fastest variability signatures \citep[see][]{raiteri2021,raiteri2021b}. Therefore, the timescales revealed by this analysis refer to the longest variability timescales present in the emission, dominating over faster variations. This is also visible through the location of the observed significant variability, appearing during the first part of the light curve, coinciding with the flaring period, when the amplitude of the variations is larger. 

\begin{figure*}
    \subfigure{\includegraphics[width=0.49\textwidth]{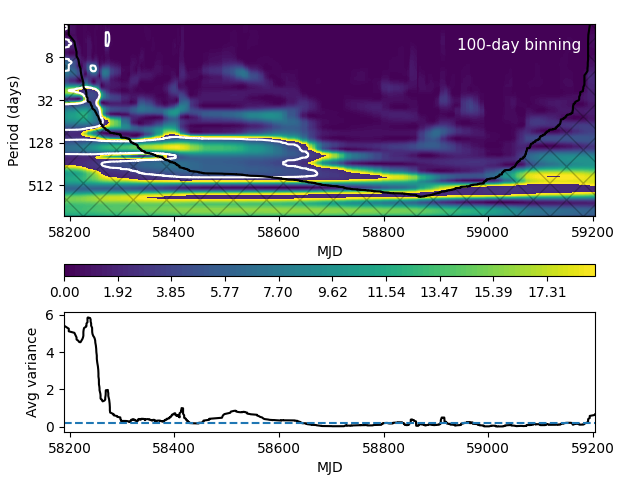}}
    \subfigure{\includegraphics[width=0.49\textwidth]{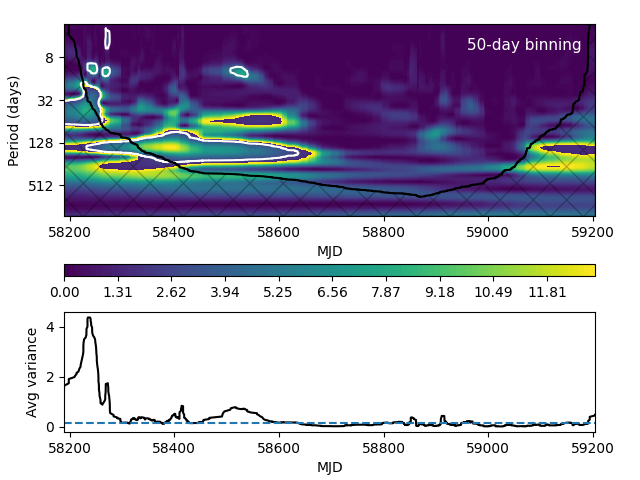}}
    \subfigure{\includegraphics[width=0.49\textwidth]{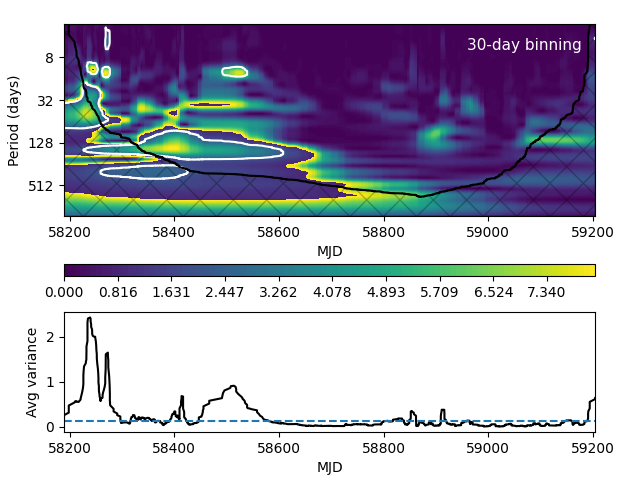}}
    \subfigure{\includegraphics[width=0.49\textwidth]{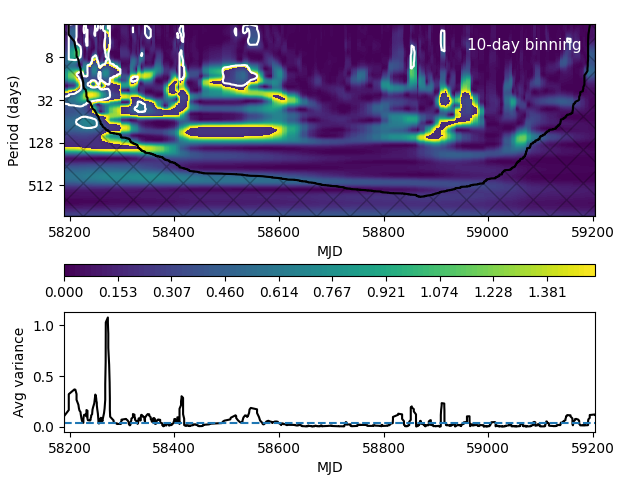}}
    \subfigure{\includegraphics[width=0.49\textwidth]{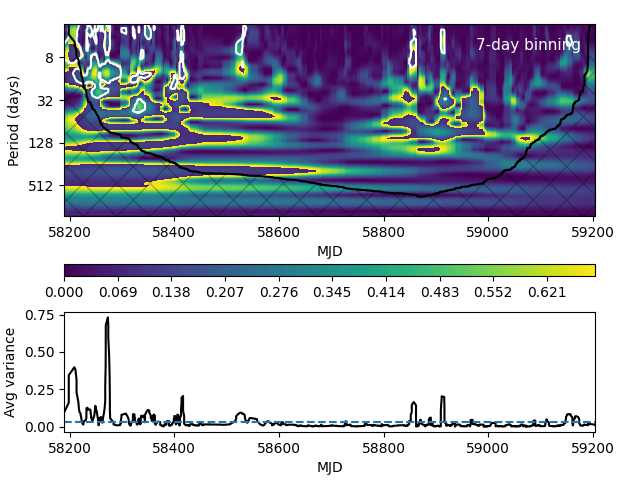}}
    \subfigure{\includegraphics[width=0.49\textwidth]{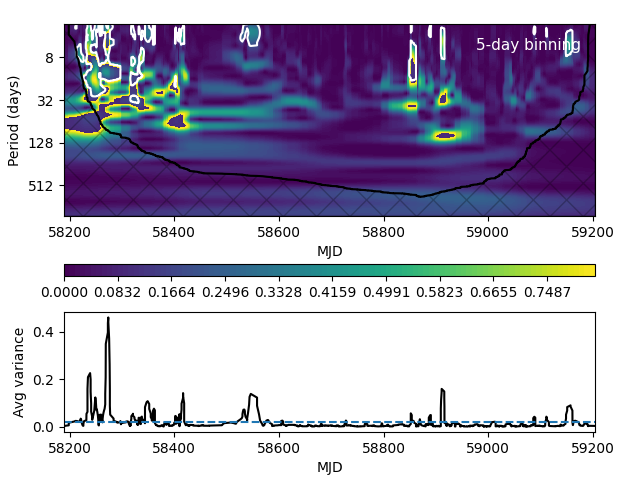}}
    \caption{Wavelet analysis performed on the detrended $R$-band light curves. \textit{Top left}: 100 day binning. \textit{Top right}: 50 day binning. \textit{Middle left}: 30 day binning. \textit{Middle right}: 10 day binning. \textit{Bottom left}: 7 day binning. \textit{Bottom right}: 5aday binning. The top panels correspond to the time-frequency power spectra of each light curve. White contours represent variability signatures with a significance of >$3\sigma$ (>99.73\% confidence level). The gray grid represents the cone of influence region affected by edge effects. The bottom panels represent the averaged variance between 1 day and the corresponding binning. The horizontal dashed line represents the 3$\sigma$ (99.73\%) confidence level.} 
    \label{fig:wavelet_LTV_detrended}
\end{figure*}

In order to unveil the effect of variability on shorter timescales we performed a detrending of the long-term $R$-band light curve following again the procedure employed by \cite{raiteri2021,raiteri2021b}. 
Owing to the timescales revealed by the wavelet analysis performed on the unbinned, observed light curve (between $\sim$100 and $\sim$300~days), we perform the cubic spline interpolation and detrending using bin widths of 100, 80, 60, 50, 40, 30, 20, 15, 10, 7, 6, 5 and 3~days. For bins smaller than 3~days, the average uncertainties of the data are larger than the standard deviation of the detrended light curve. Therefore, we cannot claim statistically significant variations below $\sim$3~days with this approach.

Figure~\ref{fig:wavelet_LTV_detrended} shows the time-period space power spectrum of the detrended light curves estimated from different bin widths. It becomes clear by visual inspection the masking effect of long timescales, as shorter significant variability signatures are revealed in the wavelet analysis when shorter temporal bins are used for obtaining the detrended light curve. Also, as the binning used for the detrending procedure becomes smaller, the features of the original light curve are better represented and thus, slower variability signatures are removed. This can be seen from the white contours representing the detected periods of significant variation. For the largest bins (100~days and 50~days), most of the variability appears with timescales >50~days. On the other hand, as we go to smaller bins, the slowest variations are more accurately removed and most of the significant variations appear at timescales <30~days. After removing the slowest variations (see binnings <10~days), several variability timescales are identified. The most remarkable ones are three $\sim$10-15-day signatures corresponding with the three flares occurring on MJD~58250, MJD~58400, and MJD~58500, approximately, as well as several characteristic timescales of $\sim$2-3 days, mainly concentrated between MJD~58200 and MJD~58600 associated with the flaring period, but also visible during the smaller flare on MJD~58900. On longer timescales, a rather stable timescale appears at $\sim$120 days, which is representative of the whole flaring period, as well as a signature at $\sim$50 days associated with the rise of the first flare covered by our observations.

We observe that the period of greatest variability occurs approximately between MJD~58200 and MJD~58600, coincident with the intense flaring period displayed by 3C~371 (see Fig.~\ref{fig:webt_LCs}). Two periods of almost no significant variability are identified. The first one spans from MJD~58600 and MJD~58800, while the second one starts roughly on MJD~59000 and lasts until the end of the observing period. Finally, in line with the red-noise-like PSD observed for the different TESS sectors, both the amplitude of the averaged variance and the power of the variability decrease as we go to shorter timescales.
%Finally, as deduced from the decreasing amplitude of the averaged variance, as well as the power of the variability, the amplitude of the variations clearly decreases as we go to shorter timescales. This is in line with the results of the variability analyses performed in the data taken by TESS and that gathered by the WEBT Collaboration, where a much higher variability amplitude was derived for the latter.

\subsection{PSD and periodicities}
\begin{figure*}
\centering
    \subfigure{\includegraphics[width=0.35\textwidth]{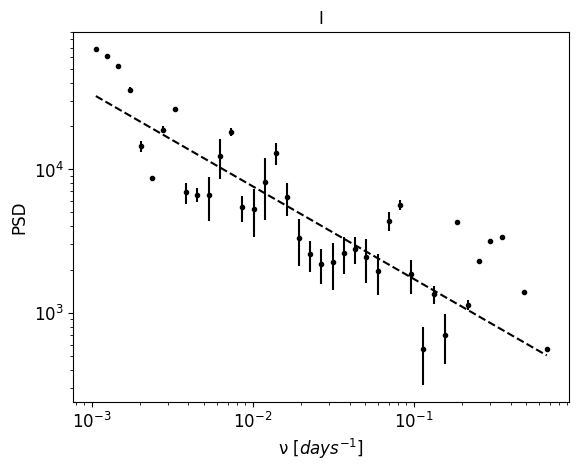}}
    \subfigure{\includegraphics[width=0.35\textwidth]{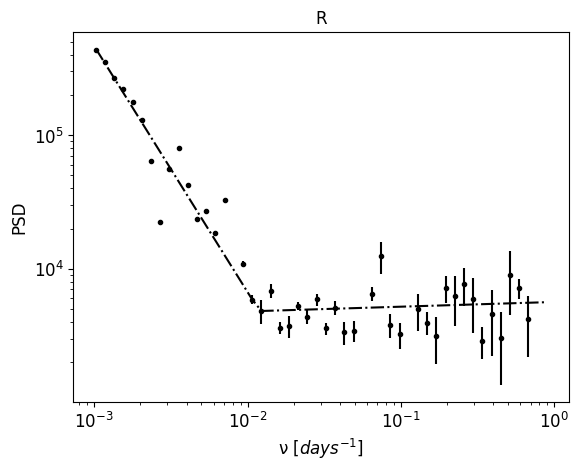}} \\
    \subfigure{\includegraphics[width=0.35\textwidth]{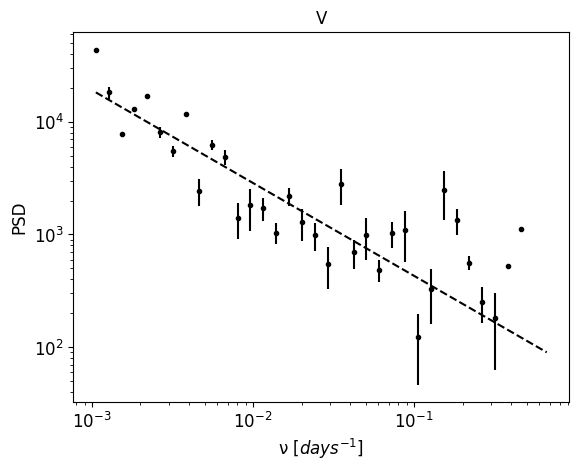}}
    \subfigure{\includegraphics[width=0.35\textwidth]{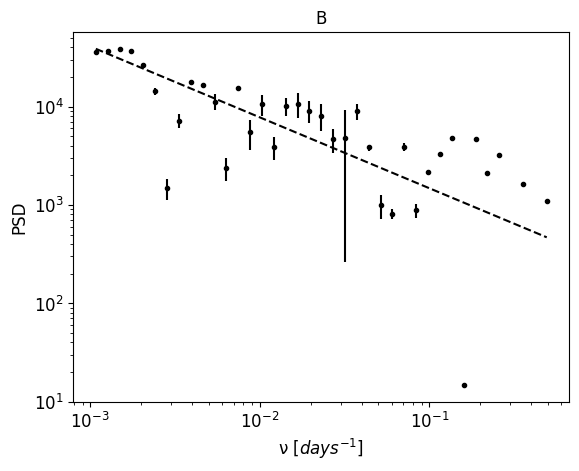}}
    \caption{Binned PSDs of the different optical bands. The dashed black lines represent the simple or broken power-law spectral shapes used for modeling the PSDs reported in Table \ref{tab:psd_mwl}.} 
    \label{fig:psd_mwl}
\end{figure*}

Similarly to the PSD analysis carried out with the data from TESS, we evaluated the PSD corresponding to the LTV of 3C~371 in its optical emission. In Fig.~\ref{fig:psd_mwl} we present the PSD of each band. The best-fit parameters are reported in Table \ref{tab:psd_mwl}. Owing to the much better sampling of the optical $R$ band, we were able to obtain a more precise characterization of the PSD in this band and thus, we were able to model it with a broken power-law function as expressed in Eq.~(\ref{eq:broken_power_law}). On the other hand, due to the more sparse sampling of the $IVB$ bands, we used a simple power law function $\mathcal{P} = A \nu^{-\alpha}$ to describe their corresponding PSDs. 
%As can be observed from these figures, there is no evident break at high frequencies, as it happened for the IDV analysis. 

\begin{table}
\centering
\caption{Results of the PSD analysis performed on the long-term optical emission of 3C~371.}
\label{tab:psd_mwl}
\resizebox{\columnwidth}{!}{%
\begin{tabular}{ccccc}
\hline
Band & A & $\alpha_1$ & $\alpha_2$ & $\nu_{b}$ [days$^{-1}$] \\ \hline
$R$  & $(4.83 \pm 2.09) \times 10^{3}$ & $1.84 \pm 0.07$ & $-0.04 \pm 0.33$ &  $0.012 \pm 0.001$  \\ \hline
$B$  & $281.21 \pm 57.81$ & $0.72 \pm 0.09$ & -- & --   \\ \hline
$V$  & $64.62 \pm 23.49$ & $0.82 \pm 0.15$ & --  & --    \\ \hline
$I$  & $390.88 \pm 49.91$ & $0.64 \pm 0.11$ & -- &  --  \\ \hline
\end{tabular}
}
\end{table}

We observe that the $R$-band PSD shows a behavior similar to that shown by the PSDs corresponding to the TESS data. The variability is represented by a power law with index $\alpha_1=1.84$, close to a red-noise like spectrum, on small frequencies. A break appears at $\nu_{b}=0.012$~days$^{-1}$, where the spectrum changes to a white-noise {dominated} regime. For the other three bands we use a power law function to model the PSD. Interestingly, these three PSDs show a smaller index, much closer to the index expected from flicker-noise like power spectrum ($\alpha \sim 1$). Flicker noise signatures in blazar PSDs have been associated with long-term variability processes with a certain ``memory'' of the variations taking place in the accretion disk, that is, disk variations have an imprint in the jet variability \citep[see for instance][]{ryan2019,bhatta2020}. However, these three bands differ from the behavior of the $R$-band PSD, with a better sampling and coverage. Therefore, no strong conclusion can be extracted in this regard with the current data.

Moreover, low-frequency breaks in the PSD of blazars have been detected in several occasions \citep[e.g.,][]{ryan2019}. These breaks have been interpreted as characteristic timescales related with the accretion disk (for instance, diffusive timescale in the outer region of the accretion flow or dynamical or thermal timescales driven by turbulence). Therefore, important information can be extracted from these breaks at low frequencies. However, no evident breaks are observed in the PSDs presented here for our data.

\begin{figure}
        \includegraphics[width=\columnwidth]{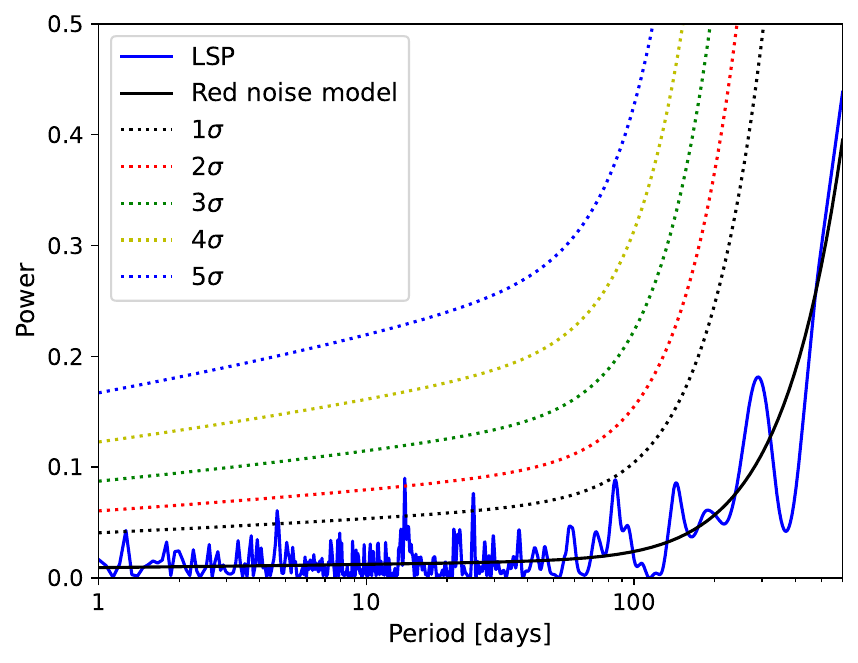}
    \caption{LSP of the $R$-band light curve of 3C~371. Blue and black solid lines correspond to the LSP and the red-noise model used for the significance estimation, respectively. Colored dotted lines represent the different confidence levels between 1$\sigma$ and 5$\sigma$.}
    \label{fig:ls_periodogram}
\end{figure}

Following the analysis of the power spectrum, we also conducted a periodicity search based on the peaks present in the periodograms/PSDs derived. Having a dataset as regularly sampled as possible is of great importance when performing a periodicity analysis, especially with Fourier transform based techniques \citep[see][]{otero-santos2023}. Therefore, we have conducted this search with the $R$-band data, for the aforementioned reasons. 
%Owing to the correlated nature of the emission in the different optical bands, if a significant periodic signature is observed in the $R$-band light curve, it is expected to be present in the rest of the optical filters. 
We employed the widely used Lomb-Scargle periodogram \citep[LSP, see][]{lomb1976,scargle1982} due to its performance on unevenly sampled data \citep[][]{penil2020,raiteri2021,raiteri2021b,otero-santos2020,otero-santos2023}. The significance estimation was performed as detailed in \cite{otero-santos2023}. In this calculation, trial factors due to the number of independent frequencies sampled have been taken into account following \cite{vaughan2005}. The resulting LSP is shown in Fig.~\ref{fig:ls_periodogram}.

We observe that the LSP presents several peaks, associated with the different characteristic variability timescales displayed by this source. However, none of these peaks have enough statistical significance to claim a possible (quasi-)periodic variation of the optical emission of 3C~371, with the highest peak barely reaching the 2$\sigma$ confidence limit. Therefore, no signs of periodicities are revealed by our analysis. This is in line with the results presented by \cite{kelly2003} and \cite{rani2009} in the radio and X-ray bands, respectively.

%\section{Spectral Variability}\label{sec5}
\section{Optical color variations}\label{sec5}

\begin{figure*}
    \subfigure{\includegraphics[width=0.67\columnwidth]{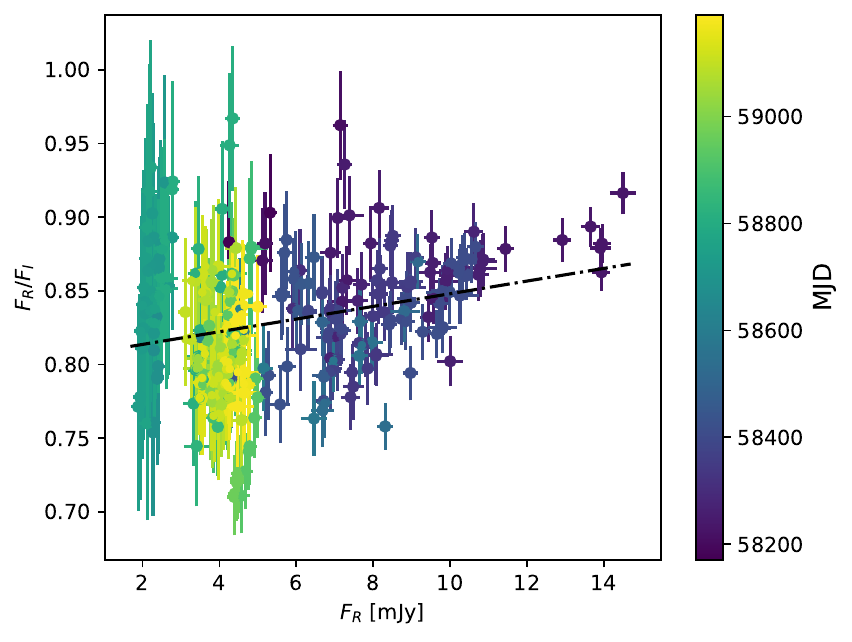}}
    \subfigure{\includegraphics[width=0.67\columnwidth]{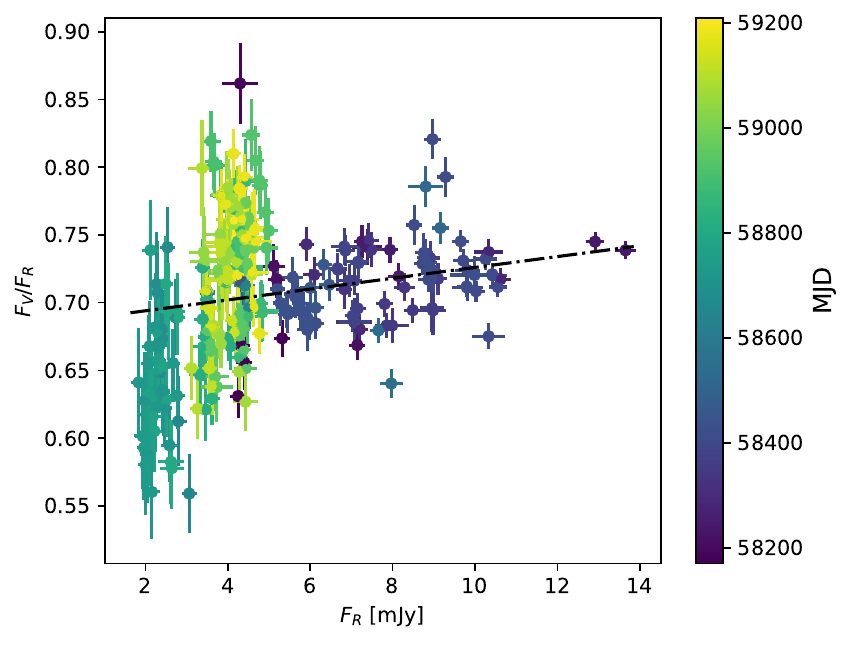}}
    \subfigure{\includegraphics[width=0.67\columnwidth]{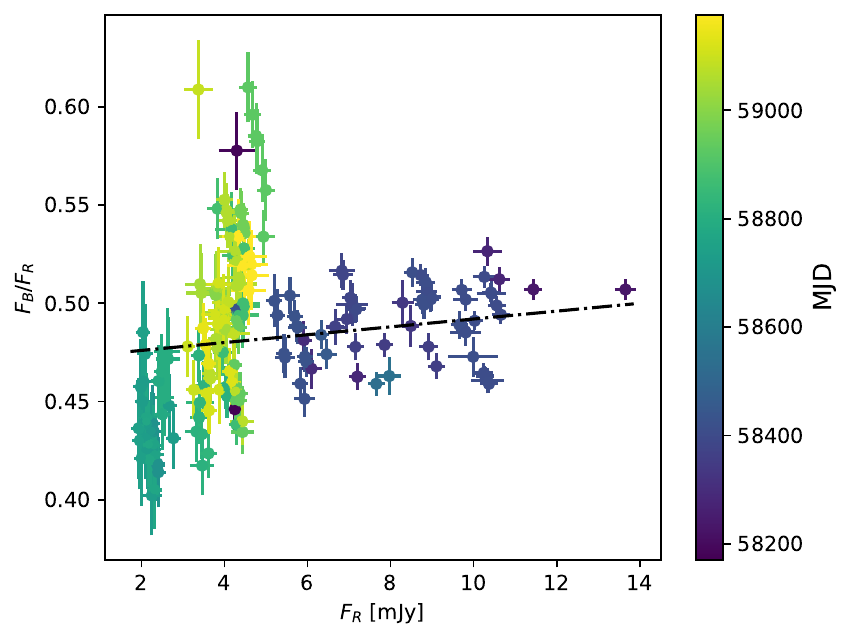}}
    \caption{Dereddened, host-galaxy-corrected color indices of 3C~371 for the different optical bands with respect to the optical $R$-band flux density. The colour scales represent the MJD of the observations. Black dashed-dotted lines correspond to the linear fits describing the color changes. \textit{Left:} $F_{R}/F_{I}$ color index. \textit{Middle:} $F_{V}/F_{R}$ color index. \textit{Right:} $F_{B}/F_{R}$ color index.} 
    \label{fig:colour_analysis}
\end{figure*}

%\subsection{Optical Colour Variations}
In order to study and understand the spectral behavior of 3C~371, we analyzed the color variations between the different optical bands over time. For this color analysis, owing to the different sampling between each band, we select the quasi-simultaneous data taken with the different filters within a range of $\pm$1~hour. All the color relations are estimated {taking the optical $R$ band as the reference because it has the largest number} of observations available. This analysis has been performed on the dereddened, host-galaxy corrected, optical flux densities. In this study, we compute the color indices as the ratios between flux density in each band and the flux density in the optical $R$ band, similarly to the approach used in \cite{otero-santos2022}. The relations between the different bands are represented in Fig.~\ref{fig:colour_analysis}, and the fit parameters are reported in Table~\ref{tab:colour_analysis_results}.

%As can be clearly seen from the optical colour-magnitude representations, this source becomes bluer as its emission becomes brighter in the optical band over long timescales. This bluer-when-brighter (BWB) behaviour has been the typical pattern observed for the BL Lac blazar subtype \citep[see for instance][]{li2018,meng2018,otero-santos2022,raiteri2023}. In particular, it has been observed in this source in the past by other authors \citep[e.g.][]{negi2022}. This behaviour can be understood as a synchrotron-dominated optical emission, as expected from BL Lac objects.

As can be seen from the optical color-magnitude representations and the coefficients of the linear fits shown in Table~\ref{tab:colour_analysis_results}, with values relatively close to zero, this source shows a rather achromatic evolution over long timescales. This behavior has been observed in the past for the LTV of other BL Lac objects \citep[e.g., S4~0954+65, S5~0716+714 or BL Lacertae, see][]{raiteri2021,raiteri2021b,raiteri2023a}. These authors interpreted this behavior as changes in the Doppler factor introduced to orientation changes of the emitting region under a helical jet model. 
%In particular, it has been observed in this source in the past by other authors \citep[e.g.][]{negi2022}. This behaviour can be understood as a synchrotron-dominated optical emission, as expected from BL Lac objects.

\begin{figure}
        \includegraphics[width=\columnwidth]{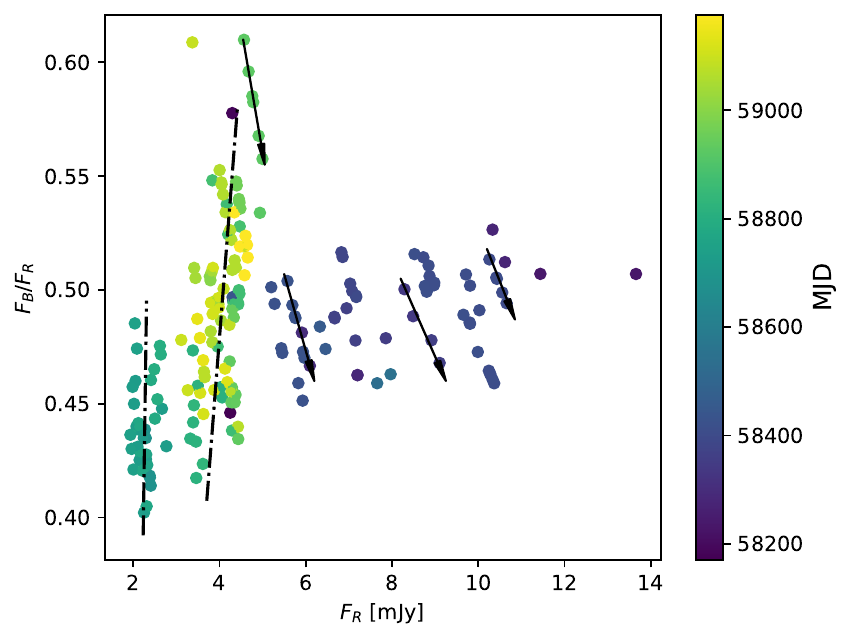}
    \caption{Dereddened, host-galaxy-corrected color index of 3C~371 for the optical $B$ band with respect to the optical $R$-band flux density. The plot is presented without error bars for clarity. The color scale represents the MJD of the observations. Black dashed-dotted lines indicate the color change of the two low states. Black arrows highlight some examples of short-term RWB variability patterns.}
    \label{fig:colour_analysis_subset}
\end{figure}

%Moreover, when considering the flaring state, we observe a rather achromatic behaviour. This lack of of a strong chromatism has been reported for other blazars in the past \citep[e.g. S4~0954+65, S50716+714 or BL Lacertae, see][]{raiteri2021,raiteri2021b,raiteri2023a} and interpreted as variability provoked by changes in the Doppler factor. 
The aforementioned approximately achromatic variability is also observed when considering only the flaring period. On the other hand, the opposite is observed during the two periods of low flux variability ($\sim$2~mJy and $\sim$4~mJy), associated however to a largely chromatic variation of the optical colors. These periods are coincident with the two low emission states of 3C~371 (dotted-dashed lines in Fig.~\ref{fig:colour_analysis_subset}, MJD~58650--58800 and MJD > 59000, approximately). Similar behaviors of very large color variation for a very small flux variability have been reported in the past for other blazars such as Mrk~421 during a high state; and 3C~454.3 or CTA~102 during a low state \citep[see][]{otero-santos2022}. 
%This clear change to a much steeper BWB slope can be indicating the existence of different electron populations producing the emission and variability, or can be associated with different emission states of the source \citep{xiong2016}.

\begin{table}
\centering
\caption{Results of the linear fits to the colour index evolution for the different bands.}
\label{tab:colour_analysis_results}
\begin{tabular}{cccc}
\hline
\multirow{2}{*}{Colour} & \multirow{2}{*}{$N$} & \multicolumn{2}{c}{Linear fit $y=a+b \cdot x$}  \\ \cline{3-4} 
  &     &    $a$     &     $b$     \\ \hline
$I-R$    &  315 &   $0.805 \pm 0.006$      &     $0.004 \pm 0.001$   \\ \hline
$V-R$   &  273  &      $0.686 \pm 0.006$      &    $0.004 \pm 0.002$     \\ \hline
$B-R$   &  196  &     $0.472 \pm 0.007$      &   $ 0.002  \pm 0.001$    \\ \hline
\end{tabular}
\end{table}

In contrast, short-term color changes display a pronounced chromatic behavior, as can be clearly seen in Fig.~\ref{fig:colour_analysis_subset}, where the $F_{B}/F_{R}$ color index is represented as an example without error bars for clarity. More remarkably, this short-term color variability displays a variety of behaviors rather than just the commonly claimed bluer-when-righter (BWB) trend for BL Lac objects. Therefore, mixed BWB and redder-when-brighter (RWB) trends occurring on short timescales can also be identified in the optical regime as signatures of fast variability. The existence of a long-term BWB variability in the color variation, accompanied with mixed BWB and RWB patterns on short timescales is an indication that LTV and STV signatures may be produced or driven by different physical processes \citep{isler2017}. For instance, \cite{negi2022} interpret these partial shorter trends as transitions between jet-dominated and disk-dominated states. The rather achromatic flaring state could indicate changes of Doppler factor \citep{raiteri2021,raiteri2021b}. Finally, long-term BWB trends are mainly expected from variations intrinsic to the synchrotron emission of the jet. 

\section{Optical polarimetry}\label{sec_polarization}
\begin{figure}
        \includegraphics[width=\columnwidth]{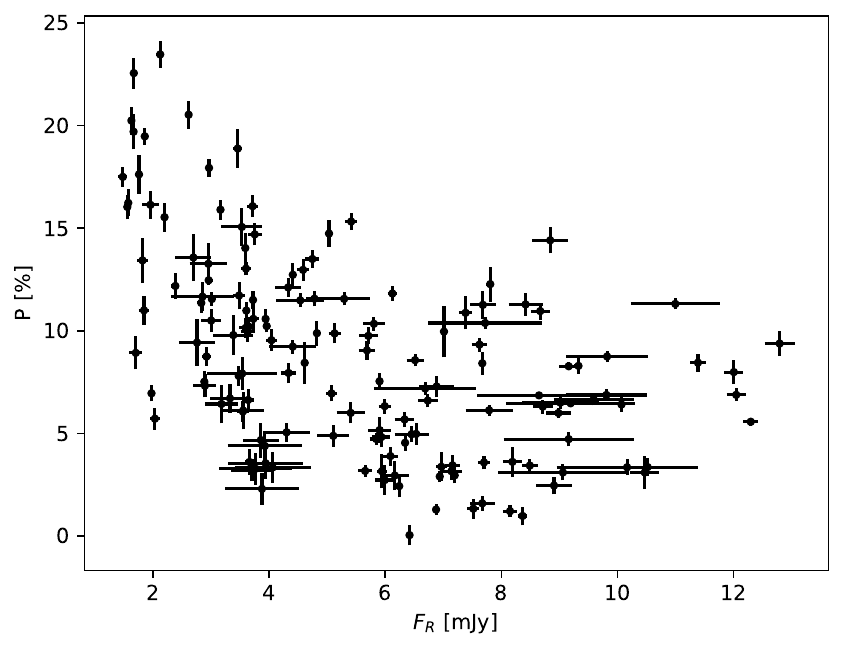}
    \caption{Intrinsic polarization degree after correcting by the depolarizing effect introduced by the host galaxy versus the derredened, host galaxy corrected, optical flux density.}
    \label{fig:polarization_degree_flux}
\end{figure}

We evaluated the behavior of the polarized emission observed from 3C~371 as part of the WEBT monitoring. First, in order to calculate the polarization fraction intrinsic to the jet, we corrected the data from the depolarizing effect introduced by the host galaxy following Eq.~(\ref{eq:depolarizing_effect}). After accounting for this effect, we observe an increase of the polarization degree up to a factor $\sim$3 during the periods with the faintest emission, when the relative contribution of the host galaxy is highest. The intrinsic polarization degree ranges from 0.05\% to 23.46\%, with an average value of 8.78\%. This results in a variability amplitude $A_{mp}=(266.52 \pm 104.72)$\%, consistent with the variability amplitude observed from the different optical bands. 
We also evaluated the possible correlation between the variability of the optical emission and the polarization degree of the jet. This can be seen in Fig.~\ref{fig:polarization_degree_flux}, where the polarization degree is represented with respect to the optical flux. 
We estimated the correlation coefficient between $P$ and $F_{R}$, resulting in a value of $\rho=-0.49$ and $\text{p-value} = 10^{-10}$, suggesting a mild anticorrelation. \cite{rajput2022} observe this behavior for three blazars on long timescales, all of them BL Lac objects. Other examples of this are 3C~345 \citep{otero-santos2023b}, BL~Lacertae \citep{raiteri2013,raiteri2023a}, and S4~0954+65 \citep{raiteri2023b}. In fact, \cite{raiteri2013} explain such behavior as a cause of a lower Lorentz factor in BL Lacs with respect to FSRQs.

%Concerning the behaviour of the EVPA, we observe two different regimes that can be clearly identified in Fig.~\ref{fig:webt_polarization_LCs}. During the intense flaring period between MJD~58100 and MJD~58600, we see a regime of large variability of the EVPA, 

Concerning the behavior of the EVPA, the angle shows an $\sim$70$^{\circ}$ alignment at the beginning of the observing period (see Fig.~\ref{fig:webt_polarization_LCs}). At the end of the first flare, a period of large variation of the angle orientation takes place, between MJD~58200 and MJD~58600 approximately, were a total angle variation of $\sim$560$^{\circ}$ occurs, from $-37^{\circ}$ to $523^{\circ}$. At the end of this interval, the EVPA shows a rather stable orientation at around 430$^{\circ}$, which corresponds to an effective EVPA change of 360$^{\circ}$ ($430^{\circ}=70+2\times 180^{\circ}$), with the angle going back to its initial orientation.

We studied this interesting feature of the EVPA evolution during the flaring period in detail. Starting on MJD~58230, 
%right after the first smooth EVPA rotation, 
the source underwent a long EVPA swing that lasted for roughly 400~days. This event showed a total effective rotation of $\sim$360$^{\circ}$, with a remarkable angle variation that ended with the EVPA aligned with the same preferential orientation observed before this event. In fact, BL Lac objects are known for showing relatively stable and preferred orientations of the EVPA \citep[see e.g.,][]{angel1980}. Therefore, this could also be the case of 3C~371, with an intrinsic process temporally disturbing the magnetic field and giving place to this feature before adopting again its original orientation. Similar features have been detected in the past for other blazars such as 3C~279, where two of these events were characterized by \cite{kiehlmann2016}, one during a very low and stable emission state and a second one during an optical flare. 

Among other recent works, \cite{kiehlmann2016} have investigated the possibility of this kind of events happening by chance, due to the stochastic variability of the emission, rather than being caused by an intrinsic process taking place in the blazar. In particular, these authors show that long and slow EVPA rotations can appear through random-walk processes, although less smooth on average. In order to assess the possibility of a physical origin for this event, we followed a similar procedure to that from \cite{kiehlmann2016}. This involved defining a smoothness parameter, $s$, for the potential rotation \citep[see Eq.~(4) in][for more details]{kiehlmann2016}, where a curve 1 is considered smoother than a curve 2 if $s_2>s_1$. With this definition we can assess the smoothness of the observed feature over similar artificially simulated EVPA rotations under a considered model. Therefore, we calculated the smoothness parameter on the daily binned, corrected EVPA values to avoid biasing $s$ with short timescale changes, in which the errors are expected to be larger than the real variation \citep[see][for further discussion]{kiehlmann2016}. The derived value of the smoothness parameter is $s=11.6$. Then, using artificial EVPA light curves obtained through MC simulations with the approach presented in Sect.~\ref{sec4}, we estimated the smoothness parameter for $10^{4}$ data series with random-walk, red-noise-like variability, with only 0.49\% of the simulations showing an EVPA behavior with the same characteristics as the real data and a smaller smoothness parameter. We performed this test for the simulations before and after the $\pm$180$^{\circ}$ correction to account for the possibility that this event is artificially introduced by an overcorrection of the EVPA. 

We note that our MC simulations are based entirely on the information of the sampling and values of the EVPA. More complex simulations are needed to confirm this result more significantly such as the ones performed by \cite{kiehlmann2016}, with information on the Stokes parameters; these allow us to constrain not only the behavior of the simulated EVPA, but also the polarization fraction and variability within different random walk theoretical models. Nevertheless, we are able to exclude at almost a 3$\sigma$ confidence level a stochastic behavior with respect to our simple stochastic model. 

Several models have been proposed for explaining such features. \cite{kiehlmann2016} comment on the possibility that a similar feature identified in 3C~279 could be produced by moving plasma along a helical magnetic field, encountering some kind of disturbance such as a shock. In fact, helical magnetic fields have successfully reproduced the long-term variation of the polarization degree and EVPA of blazars in the past \citep[see, for instance,][]{raiteri2013,raiteri2021b,raiteri2023b}.

\section{Flux variability distribution and its origin}\label{sec7}
Finally, we tested the nature of the flux distribution. Despite the common assumption of a Gaussian distribution of blazar fluxes, several authors have claimed that blazars data series are better described by a log-normal distribution \citep[see for instance][]{sinha2018,shah2018,shah2020}. This could be indicative of a nonlinear multiplicative origin of the variability, possibly linked to an imprint of the accretion disc in the jet emission, pointing towards a disk-jet connection. Nevertheless, other interpretations have also been considered in the literature. In order to investigate possible differences between the IDV and LTV signatures and origin, we evaluated the flux distribution of each day observed by TESS individually, and compared the results with the flux distribution of the 3 year $R$-band WEBT light curve.

\begin{figure}
\centering
    \subfigure{\includegraphics[width=0.95\columnwidth]{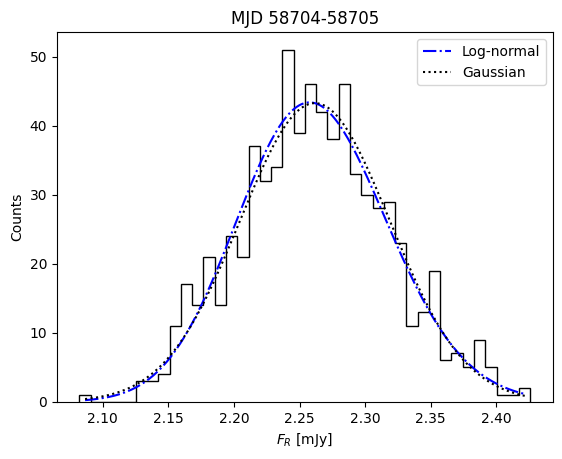}} 
    \subfigure{\includegraphics[width=0.95\columnwidth]{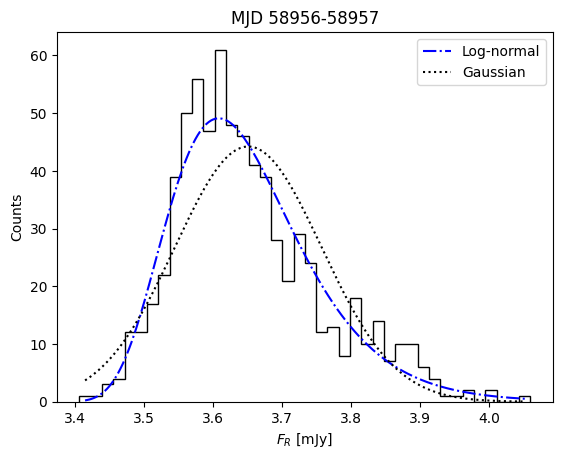}}
    \subfigure{\includegraphics[width=0.95\columnwidth]{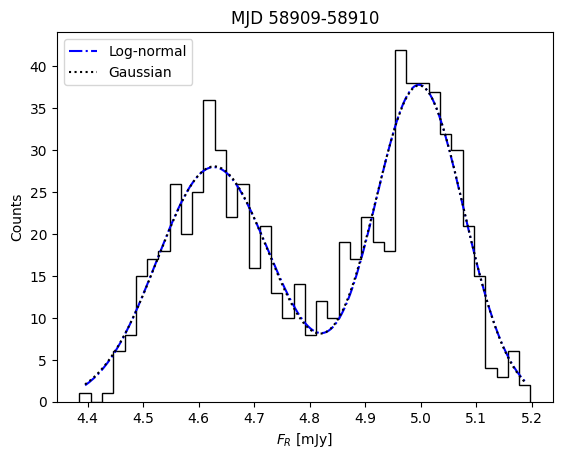}}
    \caption{Examples of IDV flux distributions for three 1 day time intervals. Black solid lines represent the histograms of the distributions. Black dotted and blue dashed-dotted lines correspond to the Gaussian and log-normal fits to the data. \textit{Top:} Flux distribution with both a compatible Gaussian and log-normal distribution. \textit{Middle.} Flux distribution with a preferential log-normal distribution. \textit{Bottom:} Multimodal flux distribution.} 
    \label{fig:IDV_flux_distributions}
\end{figure}

Most of the evaluated flux distributions for the observations performed by TESS within each 1 day interval show a good agreement with a Gaussian distribution, with values of $\chi^{2}/\text{d.o.f.} \lesssim 1$. A log-normal distribution was also tested, and was found to be also compatible with the observed PDFs, with typically no significant preference for one distribution over the other.
An example of one of the fitted distributions is shown in the bottom panel of Fig.~\ref{fig:IDV_flux_distributions}. Exceptionally, we observe that 17 out of the total 318~days show a preference for a log-normal PDF shape. Moreover, several days present multimodal distributions, which are not well-represented by either a single Gaussian nor a log-normal distribution. These time intervals are mostly coincident with the periods of fastest variability and with the most intense flares observed by TESS, especially around MJD~58810 between MJD~58850 and MJD~58920, approximately. For these cases, we find that both bi- (or multi)modal normal and log-normal PDFs are compatible with the observed distributions.

These two distributions have been associated with different origins of the observed variability. Intrinsic jet processes are known to be additive in nature and thus, leading to Gaussian PDFs \citep[see for instance][]{sinha2018}. The good performance of a Gaussian distribution in reproducing the IDV variability flux distributions may indicate that these variations are produced in a compact region of the jet through linear processes, disconnected from the disk. In fact, IDV is often interpreted within the framework of jet processes \citep[e.g., turbulence or magnetic reconnection, see][]{narayan2012}, and are difficult to explain with disk-related processes.
%, as discussed by \cite{wani2022}. 
Therefore, the results found here on the IDV flux distributions are in line with the IDV being produced by intrinsic jet processes.

\begin{figure}
        \subfigure{\includegraphics[width=\columnwidth]{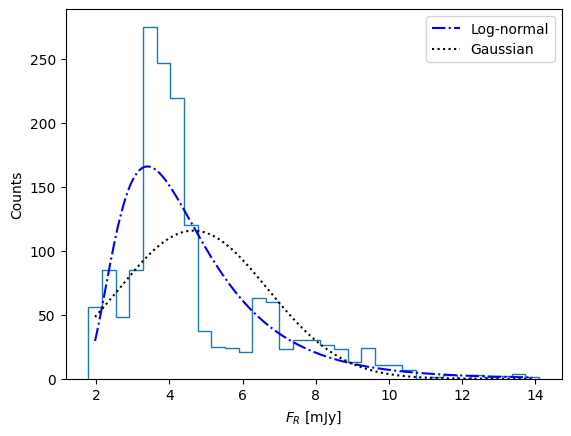}}
        \subfigure{\includegraphics[width=\columnwidth]{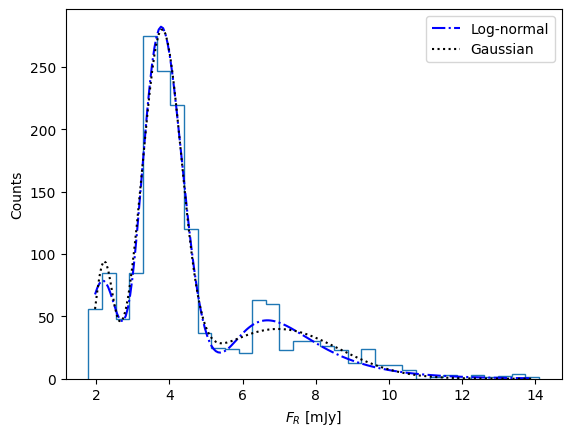}}
    \caption{Long-term flux distributions for the dereddened, host-galaxy-corrected flux density $R$-band light curve. Black solid lines represent the histograms of the distributions. Black dotted and blue dashed-dotted lines correspond to the Gaussian and log-normal fits to the data. \textit{Top:} Single log-normal and Gaussian distribution fits. \textit{Bottom:} Multimodal log-normal and Gaussian distribution fits.}
    \label{fig:LTV_flux_distributions_all}
\end{figure}

We have also studied the 3 year LTV flux distribution using the $R$-band WEBT data, as shown in Fig.~\ref{fig:LTV_flux_distributions_all}. A log-normal function represents this PDF better than a normal one. However, we observe a multimodal distribution, clearly caused by the existence of different emission states, i.e. an intense flaring period until MJD~58600 (P1); a first low-flux and relatively stable period between MJD~58600 and MJD~58820 (P2); a period showing a small flare developing between MJD~58820 and MJD~58950 (P3); and a second rather stable and low emission state starting on MJD~58950 (P4), approximately. Therefore, none of these fits results in a good agreement of the observed flux distribution, with $\chi^{2}/\text{d.o.f.}>>1$. We employed a combination of three log-normal distributions to model the PDF peaks observed at $\sim$2~mJy and $\sim$4~mJy introduced by the low states, plus the largest flux values due to the high emission state. Alternatively, we also used a composite of three Gaussian distributions. A significant improvement on the representation of the multimodal PDF can clearly be seen in the bottom panel of Fig.~\ref{fig:LTV_flux_distributions_all}, still with a preference of the log-normal distribution over the Gaussian ($\chi^{2}_{\text{log-normal}} \sim 3.6$ and $\chi^{2}_{\text{Gaussian}} \sim 47.7$).

\begin{figure*}
    \subfigure{\includegraphics[width=0.49\textwidth]{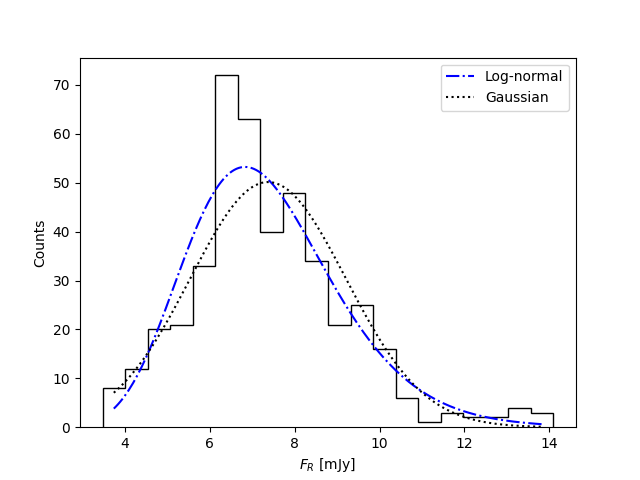}}
    \subfigure{\includegraphics[width=0.49\textwidth]{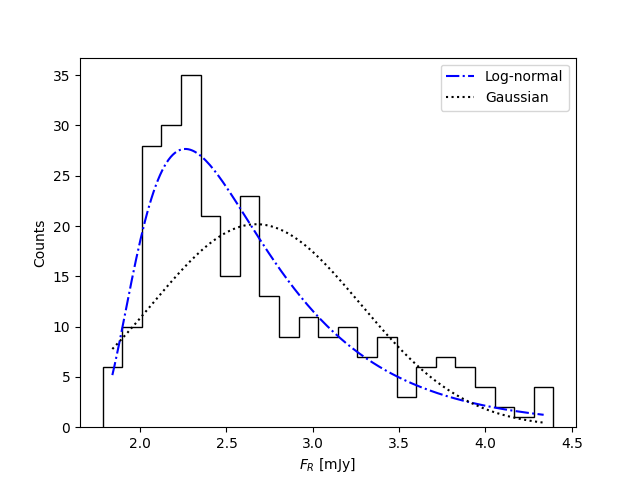}}
    \subfigure{\includegraphics[width=0.49\textwidth]{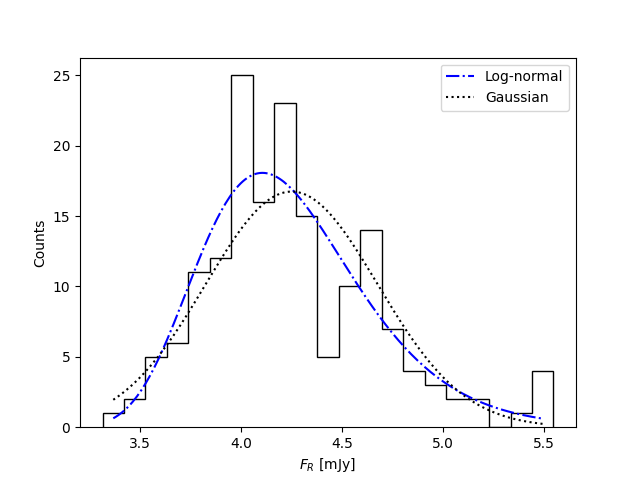}}
    \subfigure{\includegraphics[width=0.49\textwidth]{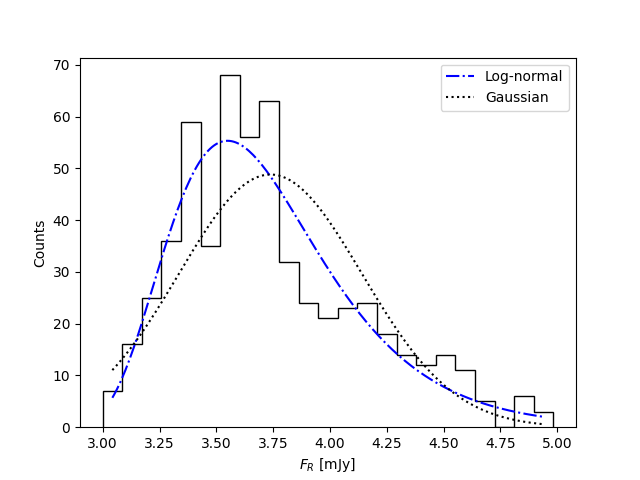}}
    \caption{Long-term flux distributions for the different emission states. Black solid lines represent the histograms of the distributions. Black dotted and blue dashed-dotted lines correspond to the Gaussian and log-normal fits to the data. \textit{Top left:} Intense flaring period (P1). \textit{Top right:} First low state period (P2). \textit{Bottom left:} Small flaring period (P3). \textit{Bottom right:} Second low state period (P4).} 
    \label{fig:LTV_flux_distributions}
\end{figure*}

Despite the better representation of the LTV flux distribution when using a multilog-normal function with respect to the single log-normal PDF, we observe from the reduced $\chi^{2}$ value that the fit still shows discrepancies with the data, especially for the data with $F_{R}>6$~mJy, i.e. during the high emission state. In order to better model the long-term flux distribution, we also evaluated the flux distribution of each period individually (see Fig.~\ref{fig:LTV_flux_distributions}). We observe a clear preference of a log-normal distribution over a normal distribution for all four periods, with $\chi^{2}_{\text{Gaussian}}>\chi^{2}_{\text{log-normal}}$. The two low-state periods and the smaller flare (P2, P3 and P4) have a reasonably good fit, with reduced $\chi^{2}$ values of 1.6, 1.9 and 2.2, respectively. The intense flaring period P1 shows a slightly higher reduced $\chi^{2}$ of 2.9, mainly due to the large number of measurements with a flux density of $\sim$6.5--7.0~mJy commented above. Nevertheless, the reduced $\chi^{2}$ value for the log-normal distribution is lower than the value obtained from the Gaussian PDF, showing a preference for the former distribution.

The prevalence of log-normal distributions has been mostly associated with processes with an origin that can be traced back to the accretion disk \citep[see e.g.,][]{kushkawa2016,bhatta2020,pininti2023}. This kind of distribution suggests that the mechanism leading to the emission and its variability is multiplicative in nature \citep{shah2020}, which is the case of accretion disk instabilities occurring at different radii in local viscous timescales, which are later propagated to the jet as multiplicative processes \citep{bhatta2020}. In our case, we observe that the long-term flux density follows a multimodal flux distribution, which has been associated in the past with different activity levels or different dominant contributing emitting regions \citep[for instance][]{pininti2023}. These could be caused by, for instance, changes in the jet orientation \citep{raiteri2021,raiteri2021b} or changes in the disk activity transmitted to the jet \citep{shah2018}. Here, different emission states can be clearly identified, leading to the observed multimodality. Evaluating the distribution of each of these emission states, we see that the LTV of each one is better represented by a log-normal distribution. Therefore, LTV may be driven by multiplicative processes, 
possibly produced in the accretion disk, later connected to the jet \citep{bhatta2020}. 
Nevertheless, there have been several claims of linear processes that could lead to non-Gaussian flux distributions. \cite{biteau2012} state that a ``minijets-in-a-jet'' model could resemble a log-normal distribution, considering observational uncertainties, invoked however to explain fast variations down to timescales of hours/minutes. Another interpretation proposed by \cite{sinha2018} involves linear fluctuations in the underlying particle acceleration and/or diffusive escape rates of the electrons, which could produce nonlinear flux perturbations, possibly explaining the observed log-normality.

\section{Summary and conclusions}\label{sec8}
In this work, we present a detailed study of the optical variability of the BL Lac object 3C~371. We investigated its properties on all timescales over which this source presents significant variability, taking advantage of the large amount of data available from the TESS satellite, with 1 year of two-minute cadence observations, and the extensive optical monitoring performed by the WEBT Collaboration. We summarize the main results in the points below:

\begin{enumerate} 
    \item We characterized the behavior of the IDV signatures displayed by the source, observing an increasing amplitude of the IDV with increasing length of the time intervals considered. We detected significant intraday flux variations with timescales ranging from several minutes to a few hours using the wavelet and ANOVA analyses, with the fastest characteristic timescales being $\tau_{\text{min}} = 0.47$~hours. This value served to constrain the size of the emitting region responsible for the observed IDV, $R \leq 6.5 \cdot 10^{14}$~cm, and the mass of the central black hole, $M_{BH} \leq 3.4 \cdot 10^{8} M_{\odot}$. We also identified the periods showing the fastest significant variations through the use of the wavelet technique, with several remarkable, very fast flaring periods.  

    \item We evaluated the PSDs for the different TESS sectors ($\tau < 28$~days) --- which were found to follow a broken power-law shape ---, revealing several IDV and STV signatures in accordance with the variability analysis. The PSD on timescales of days was found to be red-noise dominated, with $\alpha \sim 2$. Variations below the break frequency $\nu_{b}$ showed a Gaussian behavior, with the break frequencies found between $\nu_{b} \sim 5.69$~days$^{-1}$ and $\nu_{b} \sim 19.88$~days$^{-1}$. This corresponds to minimum variability timescales of between $\sim$1 and $\sim$4 hours, similar to those derived from the ANOVA and wavelet analyses. 

    \item The LTV was also characterized, showing a much larger amplitude of the variability than the IDV. In terms of IDV, we identify several characteristic variability timescales, ranging from a few days to timescales of >100~days. 

    \item We obtained the PSDs of the different optical bands. The $R$-band data, for which the best sampling was obtained, show a broken power-law shape with a red-noise like spectrum above $\nu_{b}$, typical of the LTV of blazars \citep[see e.g.,][]{raiteri2021,raiteri2021b}. %The $IVB$ bands were represented by a simple power-law function with an index closer to flicker noise, also observed in the past for some sources and associated with possible accretion disc imprints in the long-term emission \citep{bhatta2020}. 

    \item The spectral variability analysis carried out on the different optical bands shows a mild BWB behavior over long timescales; it is almost achromatic, as typically observed in the variability of BL Lac objects \citep[e.g.,][]{raiteri2021,raiteri2021b}. In addition, when evaluating shorter variations, we observe stronger and different color patters such as short BWB and RWB trends, roughly achromatic variability, or periods of large color change with almost no flux variation. 

    %\item We have constructed the optical SEDs for all the days with simultaneous $BVRI$ observations, which allowed us to identify the frequency of the synchrotron peak in the optical regime, close to the $R$-band effective wavelength. We have also estimated the optical spectral index using a power law function to model the optical SEDs, revealing a harder-when-brighter patters of the optical spectrum, often observed in BL Lac objects \citep[e.g.][]{xilouris2006}.

    \item We have analyzed the variability of the polarization degree, revealing a variability amplitude comparable to that of the optical emission. A mild anticorrelation between the polarization degree and the flux is observed. We also characterized the variability of the EVPA, showing a slow and large EVPA rotation during the flaring period. We ran simulations to decipher the probability of this feature having a physical origin. We are able to discard a simple random walk origin of the EVPA variation at a 3$\sigma$ confidence level based on these simulations.

    \item Finally, we evaluated the flux distributions over different variability timescales. We find that on timescales of 1~day, the flux is well represented by a Gaussian PDF, which could indicate that the IDV is produced in a compact region of the jet through additive processes \citep[see][]{shah2018}. On the other hand, a normal function fails to model the long-term flux distribution, which was observed to be better represented by a multilog-normal PDF. This could be pointing towards a disk-jet connection and a variability produced by multiplicative mechanisms \citep{kushkawa2016}, where the multimodal nature is due to the presence of different activity levels.
\end{enumerate}

With these results we characterize the variability of the optical emission in 3C~371 at different timescales from intraday (IDV) to short-term (STV) and long-term (LTV) scales. From the observed distributions  we derive information about its possible origin: The IDV is in agreement with Gaussian additive processes in a compact region of the jet. LTV is consistent with changes of the Doppler factor, possibly due to orientation changes of the jet, as found in previous works. However the STV shows a more stochastic behavior that may be related to intrinsic jet processes.

\begin{acknowledgements}
J.O.S. acknowledges financial support through the Severo Ochoa grant CEX2021-001131-S funded by MCIN/AEI/ 10.13039/501100011033 and through grants PID2019-107847RB-C44 and PID2022-139117NB-C44; and through grant FPI-SO from the Spanish Ministry of Economy and Competitiveness (MINECO) (research project SEV-2015-0548-17-3 and predoctoral contract BES-2017-082171).
C.M.R., M.V. and M.I.C. acknowledge financial support from the INAF Fundamental Research Funding Call 2023.
J.A.P. acknowledges financial support from the Spanish Ministry of Science and Innovation (MICINN) through the Spanish State Research Agency, under Severo Ochoa Centres of Excellence Programme 2020-2024 (CEX2019-000920-S).
C.M.R. acknowledges support from the Fundación Jesús Serra and the Instituto de Astrofísica de Canarias under the Visiting Researcher Programme 2022-2024 agreed between both institutions.
Partly based on data collected by the WEBT collaboration, which are stored in the WEBT archive at the Osservatorio Astrofisico di Torino - INAF (http://www.oato.inaf.it/blazars/webt/); for questions regarding their availability, please contact the WEBT President Massimo Villata ({\tt massimo.villata@inaf.it}).
This article is partly based on observations made with the IAC80 operated on the island of Tenerife by the Instituto de Astrofísica de Canarias in the Spanish Observatorio del Teide. Many thanks are due to the IAC support astronomers and telescope operators for supporting the observations at the IAC80 telescope. This article is also based partly on data obtained with the STELLA robotic telescopes in Tenerife, an AIP facility jointly operated by AIP and IAC.
M.D.J. thanks the Brigham Young University Department of Physics and Astronomy for continued support of the ongoing extragalactic monitoring program at the West Mountain Observatory.
This research was partially supported by the Bulgarian National Science Fund of the Ministry of Education and Science under
grant KP-06-H68/4 (2022) and by the Ministry of Education and Science of Bulgaria (support for the Bulgarian National Roadmap for Research Infrastructure).
The data presented here were obtained in part with ALFOSC, which is 
provided by the Instituto de Astrofisica de Andalucia (IAA) under a joint 
agreement with the University of Copenhagen and NOT.
E.B. acknowledges support from DGAPA-PAPIIT GRANT IN119123.
This work is partly based upon observations carried out at the Observatorio Astronómico Nacional on the Sierra San Pedro Mártir (OAN-SPM), Baja California, Mexico.
G.D., O.V. and M.S. acknowledge support by the Astronomical Station Vidojevica, funding from the Ministry of Science, Technological Development and Innovation of the Republic of Serbia (contract No. 451-03-47/2023-01/200002), by the EC through project BELISSIMA (call FP7-REGPOT-2010-5, No. 256772), the observing and financial grant support from the Institute of Astronomy and Rozhen NAO BAS through the bilateral SANU-BAN joint research project "GAIA astrometry and fast variable astronomical objects", and support by the SANU project F-187.
This work is partly based upon observations carried out at the Hans Haffner Observatory.
This paper includes data collected by the TESS mission, which are publicly available from the Mikulski Archive for Space Telescopes (MAST).
Python wavelet software provided by Evgeniya Predybaylo based on \cite{torrence1998} and is available at URL: \url{http://atoc.colorado.edu/research/wavelets/}.
The authors thank the referee for the very careful review of the manuscript and the useful comments.
\end{acknowledgements}

   \bibliographystyle{aa} % style aa.bst
   \bibliography{biblio} % your references Yourfile.bib

\end{document}